\documentclass[
    11pt,a4paper
]{article}

\usepackage{jheppub}

\pdfoutput=1
\usepackage[utf8]{inputenc}
\usepackage[T1]{fontenc}
\usepackage[svgnames]{xcolor}
\usepackage{
    amsmath,
    amssymb,
    amsfonts,
    mathrsfs,
    amsthm,
    lmodern,
    microtype,
    bm,
    epstopdf,
    graphics,
    booktabs,
    mathtools,
    subcaption,
    physics,
    verbatim,
    csquotes,
    float,
    multirow
}
\usepackage[normalem]{ulem}

\definecolor{ultramarine}{rgb}{0.07, 0.04, 0.56}
\definecolor{cadmiumgreen}{rgb}{0.0, 0.42, 0.24}
\definecolor{indigo(dye)}{rgb}{0.0, 0.25, 0.42}

\numberwithin{equation}{section}

\usepackage{array}
\newcolumntype{P}[1]{>{\centering\arraybackslash}p{#1}}

\usepackage[shortlabels]{enumitem}

\usepackage{dcolumn}
\newcolumntype{M}[1]{>{\centering\arraybackslash}m{#1}}
\newcolumntype{N}{@{}m{0pt}@{}}

\newcommand{\Mpl}{M_{\text{Pl}}}
\newcommand{\md}{\mathrm{d}}
\newcommand{\Dsp}{\mathscr D}

\title{Genesis in Minimally Modified Gravity}

\author[a]{Alexander Ganz}
\author[b]{Paul Martens}
\author[c,d,e]{Shinji Mukohyama}
\author[f]{Pavel Petrov}

\affiliation[a]{Institute for Theoretical Physics, Leibniz University Hannover, Appelstraße 2, 30167 Hannover, Germany}
\affiliation[b]{Department of Physics, The Chinese University of Hong Kong (CUHK),
Shatin, New Territories, Hong Kong}
\affiliation[c]{Research Center for the Early Universe (RESCEU), Graduate School of Science, The University of Tokyo, Hongo 7-3-1, Bunkyo-ku, Tokyo 113-0033, Japan}
\affiliation[d]{Center for Gravitational Physics and Quantum Information, Yukawa Institute for Theoretical Physics, Kyoto University, 606-8502, Kyoto, Japan}
\affiliation[e]{Kavli Institute for the Physics and Mathematics of the Universe (WPI), The University of Tokyo Institutes for Advanced Study (UTIAS), The University of Tokyo, Kashiwa, Chiba 277-8583, Japan}
\affiliation[f]{Cosmology, Gravity, and Astroparticle Physics Group, Center for Theoretical Physics of the Universe, Institute for Basic Science (IBS), Daejeon, 34126, Korea}

\emailAdd{alexander.ganz@itp.uni-hannover.de}
\emailAdd{paulmartens@cuhk.edu.hk}
\emailAdd{shinji.mukohyama@resceu.s.u-tokyo.ac.jp}
\emailAdd{p669590371@ibs.re.kr}

\abstract{
We construct a Genesis scenario within the framework of a type-II minimally modified gravity.
The background solution is past geodesically complete and approaches Minkowski spacetime in the asymptotic past.
While the Hubble expansion rate grows, the matter sector satisfies the null-energy condition.
Scalar perturbations are ghost-free and have no ultraviolet gradient instability, although the squared sound speed $c_\mathcal{R}^2$ becomes negative within a finite infrared region in the Fourier space for a finite time interval. Numerical evolution, including superhorizon growth, yields a nearly scale-invariant, slightly red-tilted scalar spectrum over a broad region of parameter space.
The tensor spectrum is blue, with a strongly suppressed tensor-to-scalar ratio.
Finally, using the matter fluctuation $\delta\chi$, whose asymptotic normalization is regular, we find that the background-dependent cubic couplings vanish in the far past at fixed nonzero momenta.
The surviving matter--graviton and graviton self-interactions are Planck suppressed.
}

\begin{document}

\begin{flushright} 
{\footnotesize RESCEU-28/26, YITP-26-145, IPMU26-0037}
\end{flushright}
\vspace{0.5cm}

\maketitle
\flushbottom

\section{Introduction}
\label{sec:Introduction}

Genesis scenarios describe a universe that approaches Minkowski spacetime
in the asymptotic past before entering an expanding phase with an
increasing Hubble expansion rate. The scale factor tends to a nonzero constant
while the Hubble expansion rate tends to zero in the infinite past, avoiding an initial curvature
singularity. The original Galilean Genesis construction demonstrated that
such an evolution can be supported by a scalar field with higher-derivative
interactions~\cite{Creminelli:2010ba,Creminelli:2012my}. During Genesis,
$\epsilon=-\dot H/H^2<0$. In a spatially flat universe governed by general
relativity (GR), this requires violating the null-energy condition
(NEC)~\cite{Creminelli:2006xe,Rubakov:2014jja}. Realizing this evolution
without pathological perturbations is therefore a central challenge.

A suitable background alone does not guarantee an acceptable primordial
power spectrum. In the original Genesis scenario, a nearly scale-invariant
spectrum can be generated by an additional spectator
field~\cite{Creminelli:2010ba,Creminelli:2012my}, while generalized Genesis
models also allow nearly scale-invariant curvature perturbations to be
generated directly~\cite{Nishi:2015pta}. The relevant observational
requirements concern the scalar amplitude, spectral index and running,
as well as the tensor spectrum and primordial non-Gaussianities. We use
the results of Planck 2018~\cite{Planck:2018jri,Planck:2019kim} as an
observational reference.
Since the curvature perturbations need not be conserved outside the horizon
during Genesis, its subsequent evolution must be included when determining
the spectrum at the end of this phase.

The stability of nonsingular cosmologies has been studied extensively in
scalar--tensor theories. No-go results in generalized Galileon and
Horndeski theories show that, under their stated assumptions, a
nonsingular background cannot remain free of instabilities throughout
its evolution~\cite{Libanov:2016kfc,Kobayashi:2016xpl}. These conclusions
depend on the asymptotic behavior of the perturbation coefficients and
on the theory under consideration. Stable Genesis solutions have, for
example, been constructed in beyond-Horndeski and DHOST
theories~\cite{Kolevatov:2017voe,Mironov:2019qjt,Ilyas:2020zcb,Zhu:2021ggm}.
Linear stability is also distinct from perturbative control: a
time-dependent strong-coupling scale must be compared with the
characteristic energy scale of the background, rather than judged from
its asymptotic value alone. Analyses at cubic and higher orders have
identified Horndeski Genesis regimes in which this hierarchy can be
maintained~\cite{Ageeva:2020gti,Ageeva:2020buc,GilChoi:2025hbs}.

In this work, we consider VCDM, a theory of type-II minimally modified
gravity~\cite{DeFelice:2020eju}. (For the relation between VCDM and cuscuton~\cite{Afshordi:2006ad}, see \cite{DeFelice:2022uxv}.) Its constraint structure leaves only two
propagating tensor degrees of freedom in the gravitational sector, at
the cost of breaking time-diffeomorphism invariance. The auxiliary scalar
introduces no additional propagating degree of freedom, while its
potential allows the background expansion to be
reconstructed~\cite{DeFelice:2020eju,DeFelice:2020cpt}. This freedom has
already been used to construct bouncing cosmologies with a canonical
matter scalar~\cite{Ganz:2022zgs}. Here we investigate whether VCDM can
support an asymptotically Minkowski Genesis phase with canonical
NEC-respecting matter, a nearly scale-invariant curvature spectrum, and
cubic interactions that remain controlled in the asymptotic past.

We construct the background using a power-law evolution of the conformal
Hubble expansion rate and a logarithmic evolution of the matter scalar. The
matter potential is obtained in terms of exponential integral functions,
while the monotonic evolution of the auxiliary scalar determines the
VCDM potential parametrically. The matter sector satisfies the null energy condition (NEC)
\begin{equation}
        \rho_\chi+p_\chi =\dot{\bar\chi}^{\,2}>0\,,
\end{equation}
while the modified gravitational equations allow $\dot{H}/H^2>0$. Thus,
the effective NEC-violating expansion is generated by the modified
gravitational dynamics, rather than by NEC violation in the matter sector.

We study the scalar and tensor perturbations using analytical
approximations and numerical evolution. The curvature perturbation has
a positive kinetic coefficient throughout Genesis, and its
short-wavelength sound speed approaches unity.
At finite momentum, $c_\mathcal{R}^2$ can become negative within a bounded
infrared region. However, since $c_\mathcal{R}^2\to1$ in the ultraviolet,
this does not lead to the arbitrarily fast short-wavelength growth
characteristic of an ultraviolet gradient instability.
After accounting for the superhorizon evolution, the numerical scalar
spectrum exhibits a nearly scale-invariant,
slightly red-tilted regime over a broad range of model parameters. The
tensor modes have the GR quadratic action, and their spectrum is blue,
with a strongly suppressed tensor-to-scalar ratio for the parameters
considered.

The cubic analysis requires particular care in the choice of perturbation
variable. Nonlinear definitions of the curvature perturbation and their
application to VCDM have been discussed in~\cite{Ganz:2024ihb}. On the
present background, the nonlinear transformation to the curvature
perturbation can become singular in the deep Genesis limit. We therefore
analyze the asymptotic interactions directly in terms of the matter
fluctuation $\delta\chi$, whose normalization remains regular in the
fixed-momentum limit. Considering the pure scalar, mixed scalar--tensor
and pure tensor cubic sectors, we find that the background-dependent
cubic couplings vanish in the asymptotic past, while the surviving
matter--graviton and graviton self-interactions remain Planck suppressed.
This result holds at fixed nonzero momenta and away from the singular surface of the $\delta\chi$ velocity reduction. Thus, the cubic power counting reveals no
strong-coupling scale that collapses to zero in this limit.
However, it does not determine the full nonlinear effective field theory cutoff.

The paper is organized as follows. Section~\ref{sec:the_model} introduces
the model and reconstructs the matter scalar potential and the VCDM potential from the Genesis background.
Section~\ref{sec:Linear_perturbations} analyzes linear perturbations,
their stability, the scalar and tensor power spectra, and past geodesic
completeness.
Section~\ref{sec:Cubic_Interaction} presents the cubic analysis.
Section~\ref{sec:Conclusion} summarizes the results.
Appendix~\ref{app:Phenomenological_fit} gives the phenomenological fit to
the scalar spectrum.
Appendix~\ref{app:Absence_Strong_Coupling} provides the reduced perturbation
actions and asymptotic cubic power counting.

\section{The VCDM model}
\label{sec:the_model}

\subsection{General description}

The VCDM construction is described in detail in~\cite{DeFelice:2020eju}.
Here, we briefly review the definitions required for a general understanding.
The VCDM action reads
\begin{equation}
    \begin{aligned}
        S_\text{VCDM} = \frac{\Mpl^2}{2} \int \md^4x N \sqrt{\gamma}
        \Bigg[
            &
                {}^{(3)} R
                + K_{ij} K^{ij} - K^2
                - 2 \, V(\varphi) \\
            &
                - 2 \, \frac{\lambda^i}{N} \, \partial_i \varphi
                - \frac{3}{2} \, \lambda_0^2
                - 2 \lambda_0 \left( K + \varphi \right)
        \Bigg]\,,
    \end{aligned}
    \label{action_mmg}
\end{equation}
where $K_{ij}$ is the extrinsic curvature, ${}^{(3)} R$ is the Ricci scalar associated with the metric $\gamma_{ij}$ on the hypersurfaces of constant time, $N$ is the lapse function, the quantities $\lambda^i$ and $\lambda_0$ are Lagrange multipliers, and $\varphi$ is an auxiliary scalar field. The nontrivial constraint structure leaves only two propagating tensor degrees of freedom, at the cost of breaking time-diffeomorphism invariance.
In addition, we consider a canonical scalar field for the matter sector
\begin{equation}
    \label{action_matter}
    S_\text{matter} =
        \int \md^4x N \sqrt{\gamma} \,
        \left(
            - \frac{1}{2} g^{\mu\nu} \partial_\mu \chi \partial_\nu \chi
            - U(\chi)
        \right)\,.
\end{equation}
The total action thus reads
\begin{equation}
  \label{action_total}
  S = S_\text{VCDM} + S_\text{matter}\,.
\end{equation}

\subsection{Genesis background}
For a homogeneous and isotropic background, we take the following background quantities
\begin{align}
  \begin{aligned}
    \label{BG_metric}
    N & = \bar{N}(t)\,,
      & N^i & = 0\,,
      & \gamma_{ij} & = a^2(t) \, \delta_{ij}\,,
      & \varphi & = \phi(t)\,, \\
     \qquad \lambda^i & = 0\,,
      & \lambda_0 & = \bar\lambda(t)\,,
      & \chi & = \bar\chi(t)\,,
  \end{aligned}
\end{align}
where $N^i$ is the shift vector. Varying the action \eqref{action_total} with respect to $\bar N, \, \bar\lambda, \, \phi, \, a, \, \bar\chi$ leads, respectively, to
\begin{subequations}
  \begin{align}
    \label{eq:3}
    3 H^2 & = V + \phi \bar \lambda + \frac{3}{4} \, \bar \lambda^2 + \frac{\dot{\bar \chi}^2}{2 \Mpl^2}  + \frac{U}{\Mpl^2}\,, \\
    0 & = 3 H + \phi + \frac{3}{2} \, \bar\lambda\,, \\
    0 & = \bar\lambda + V_\varphi\,, \\
    2 \, \frac{\partial_t H}{\bar N} + 3 H^2 & = V + \phi \bar\lambda + \frac{3}{4} \, \bar\lambda^2 - \frac{\partial_t \bar\lambda}{\bar N} - \frac{\dot{\bar \chi}^2}{2 \Mpl^2}  + \frac{U}{\Mpl^2}\,, \\
    0 & = \ddot{\bar \chi} + 3 H \dot{\bar \chi} + U_{,\chi} \,, \label{eq:EOM_matter_sector}
  \end{align}
\end{subequations}
where $H \equiv \partial_t a \cdot (aN)^{-1}$ is the Hubble expansion rate, and $V$ and $U$ (and their derivatives) are all evaluated at the background values. Dots denote proper-time derivatives, that is $\dot f \equiv \bar N^{-1}\partial_t f$, and primes denote conformal-time derivatives $\md/\md\tau \equiv a \bar N^{-1}\partial_t$.

We start from an ansatz for the conformal Hubble expansion rate and subsequently
reconstruct the corresponding VCDM potential $V$.
Working in conformal time $\tau\equiv \int^td\tilde{t}\bar{N}(\tilde{t})/a(\tilde{t})$, we take
\begin{equation}
    \mathcal{H} = \frac{h_0}{(-\tau-\tau_B)^{d+1}}\,,
    \qquad
    a = e^{\frac{h_0}{d}(-\tau-\tau_B)^{-d}}\,,
\end{equation}
where $\mathcal{H} \equiv a'/a$ is the conformal Hubble expansion rate, $\tau_B<0$ and $d>0$ are constant. 
The \enquote{slow-roll} parameter $\epsilon$ is given by
\begin{equation}
 \epsilon = - \frac{\dot H}{H^2} = 1- \frac{(1+d) (-\tau_B)^d }{h_0} (1+x)^{d}\,,
\end{equation}
where we have introduced the dimensionless time $x = \tau/\tau_B$. For later convenience, we define $T \equiv 1+x$.
We choose $h_0$ such that $\tau=x=0$ marks the end of the Genesis and that $\tau<0$ during the Genesis epoch, i.e. $\epsilon(0)=0$, which gives
\begin{equation}
    \frac{h_0}{(-\tau_B)^d}=1+d\,.
\end{equation}
The approximation $a \simeq 1$ is valid only when $(1+d)T^{-d}/d\ll1$ and
therefore breaks down near the endpoint, where
\begin{equation}
    a(0)=e^{\frac{1+d}{d}}\,.
\end{equation}
The endpoint Hubble scale is
\begin{equation}
    H(0)=(1+d) \, e^{-\frac{1+d}{d}}(-\tau_B)^{-1}\,,
\end{equation}
so at fixed $d$, it is $(-\tau_B)^{-1}$ that sets the characteristic energy scale.
We restrict our analysis to the Genesis phase.
Its transition to a subsequent inflationary or reheating stage is left for a future work.

\subsection{Evolution of the matter sector}

To keep the analysis tractable, we assume a logarithmic evolution for the background matter field:
\begin{equation}
    \label{eq:Ansatz_chi}
    \bar \chi(\tau) = \chi_1 \log \Big[ M_\star (- \tau-\tau_B) \Big] = \chi_1 \left[ \log (1 +x) + \log( - M_\star \tau_B) \right]\,,
\end{equation}
where $M_\star$ is an auxiliary mass scale that makes the logarithm dimensionless. Here we consider $M_\star>0$ and $\chi_1>0$. Note that deep in the Genesis epoch the scalar field grows without bound, $\lim_{x\rightarrow \infty} \bar\chi = \infty$. For later purposes, it is useful to introduce
\begin{equation}
    \alpha = \frac{\rho_\chi + p_\chi}{\Mpl^2 H^2} = \frac{\dot{\bar \chi}^2}{ \Mpl^2 H^2} = \frac{\chi_1^2  }{ \Mpl^2 (1+d)^2} (1+x)^{2d} \equiv \alpha_0 (1+x)^{2d}> 0\,.
\end{equation}
Unlike in GR, $\alpha$ is not directly tied to $\epsilon$.
In the far past, $\alpha\to\infty$, while
$\alpha\mathcal{H}^2=a^2(\rho_\chi+p_\chi) \, \Mpl^{-2} \to 0$.
At the endpoint $\alpha(0)=\alpha_0$, and below we focus on
$\alpha_0\ll1$.

\begin{figure}
    \centering
    \includegraphics[width=0.5\linewidth]{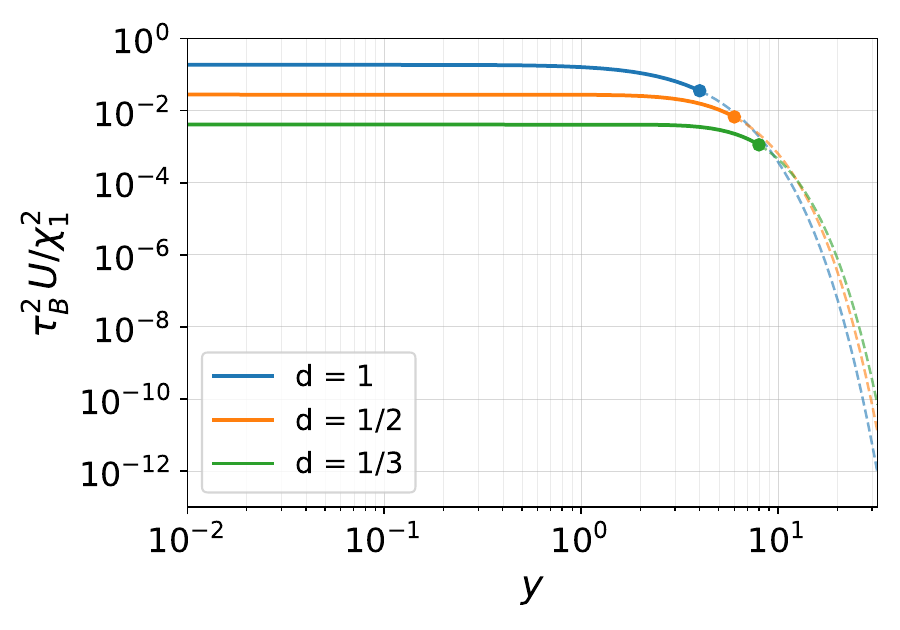}
    \caption{
Reconstructed matter potential $U(y)$ for $d=1/3$ (green), $1/2$ (orange),
and $1$ (blue). Solid curves correspond to the Genesis trajectory,
$0<y\le 2(1+d)/d$, and the markers indicate the end of Genesis at
$y=2(1+d)/d$. Dashed curves show the mathematical continuation of
$U(y)$ beyond the Genesis domain.
}
    \label{fig:Matter_Potential}
\end{figure}

Substituting the ansatz \eqref{eq:Ansatz_chi} into the field equation \eqref{eq:EOM_matter_sector}, we get an equation for the potential $U$ that reads
\begin{equation}
    -  \chi_1 (- \tau- \tau_B)^{-2} - 2 h_0 \chi_1 (- \tau -\tau_B)^{-d-2 } + a(\tau)^2   \frac{\md }{\md \chi} U =0\,.
\end{equation}
This equation can be solved exactly using $\md \chi = - \chi_1 \, (-\tau-\tau_B)^{-1} \md \tau $ as
\begin{equation}
    U(\chi) = \frac{\chi_1^2}{\tau_B^2} \left( \frac{d\,y}{2(1+d) }   \right)^{2/d} \left( \frac{1}{d} E_{1-\frac{2}{d}}\left( y\right) +  y E_{-\frac{2}{d}}(y) \right)\,,
\end{equation}
where $E_\nu(y)=\int_1^\infty e^{-yt}t^{-\nu}\md t$. The additive constant is chosen so that $U(y\to\infty)=0$, and
\begin{equation}
    y = \frac{2 (1+d)}{d} e^{- \frac{d \chi}{\chi_1}} (-\tau_B M_\star)^d\,.
\end{equation}
During the Genesis phase, $y$ is restricted to $0 < y \leq 2 (1+d)/d$. In general, the potential has essentially the form of a step function. Deep in the Genesis phase, $x\to\infty$, the scalar field lies near the
top of the step-like potential and rolls downhill with $\dot{\bar\chi}\to0$.
The parameter $d$ controls the steepness of the potential, as shown in
figure~\ref{fig:Matter_Potential}.

At the end of the Genesis phase, the energy density is given by
\begin{subequations}
    \begin{align}
        \rho_\chi(0) & = \left. \frac{1}{2} \dot{\bar \chi}^2 + U(\chi) \right|_{\tau=0} \\
        & = \frac{\chi_1^2}{ \tau_B^2} \left[ \frac{1}{2} e^{-\frac{2+2d}{d}} + \left( 2 + \frac{2}{d} \right) E_{-\frac{2}{d}} \left(
            \frac{2+2d}{d}
        \right)
        + \frac{1}{d} E_{1-\frac{2}{d}}
        \left(
            \frac{2+2d}{d}
        \right)
        \right]\,.
    \end{align}
\end{subequations}
The kinetic and potential contributions are of the same order. For the parameters used below, $\Omega_\chi(0)\equiv\rho_\chi(0)/(3\Mpl^2H(0)^2)\ll1$.

\subsection{Reconstruction of the VCDM potential}

To reproduce the chosen Hubble evolution, we reconstruct the VCDM potential following~\cite{DeFelice:2020cpt,Ganz:2022zgs}. The background equations determine the auxiliary field:
\begin{align}
    \phi & = \frac{3}{2 \Mpl^2} \int \md \tau\, a \dot{\bar \chi}^2
    = \frac{3}{2 \Mpl^2} \int \md \tau\, \frac{\bar\chi^{\prime 2}}{a} \nonumber \\
    & = -\frac{3 \chi_1^2}{2 (-\tau_B) \Mpl^2} \int \md x \frac{e^{-\frac{1+d}{d} (1 + x)^{-d}}}{(1+x)^2} \nonumber \\
    & = - \frac{3 \chi_1^2 }{2 \Mpl^2 d  (- \tau - \tau_B) } E_{\frac{d-1}{d}}\left( \frac{y}{2} \right) + \phi_0\,,
\end{align}
where we used $\dot{\bar\chi}=\bar\chi^\prime/a$ and where $\phi_0$ is an integration constant, which we will set to $\phi_0=0$. As $\phi$ is a strictly monotonic function, one can invert the relation to obtain $x=x(\phi)$. For generic $d$ the inversion is numerical only, while special cases such as $d=1$ admit a closed form. Substituting $\phi(x)$ into the background equations gives
\begin{align}
    V = & \frac{\phi^2}{3} - \frac{\rho_\chi}{\Mpl^2} \nonumber \\
    = & \frac{3 \chi_1^4}{4 \Mpl^4 d^2 (- \tau - \tau_B)^2} E_{\frac{-1+d}{d}}\left(\frac{y}{2}\right)^2 - \frac{\chi_1^2 e^{- 2\,\frac{1+d}{d} (1+x)^{-d}}}{2 \Mpl^2 (- \tau - \tau_B)^2} \nonumber \\
    & - \frac{\chi_1^2}{  \Mpl^2 \tau_B^2} \left( \frac{d\,y}{2(1+d)} \right)^{2/d} \left( \frac{1}{d} E_{1-2/d}(y) + y E_{-2/d}(y) \right)\,,
\end{align}
where $y=y(x)$ should be understood as a function of $x$.
For generic $d$, $V(\varphi)$ is reconstructed numerically.
For $0<d<1$, the potential and its first two derivatives remain regular
in the asymptotic past,
while the third derivative diverges.
This does not affect nonzero-momentum perturbations, since the
$\lambda^i$ constraint implies $\delta\varphi_{\bm k}=0$ for
$\bm k\neq0$.

\section{Linear perturbations}
\label{sec:Linear_perturbations}

The VCDM action is invariant under foliation-preserving transformations $t\to\widetilde t(t)$ and $x^i\to\widetilde x^i(t,\bm x)$, but not under spatially dependent changes of time slicing. We use the spatial gauge freedom to remove the scalar off-diagonal metric perturbation. We thus perturb the lapse, shift, and spatial metric as
\begin{equation}
  \label{eq:11}
  N = \bar{N}(t) \left( 1 + \delta N \right) \,, \qquad
  N^i = \frac{\bar{N}(t)}{a(t)} \delta^{ji} \partial_j \beta   \,, \qquad
  \gamma_{ij} = a^2(t) \, e^{2 \zeta} (e^{h})_{ij}\,,
\end{equation}
where $h_{ij}$ is transverse traceless: $\delta^{ij}\partial_i h_{jk}=0$, $\delta^{ij}h_{ij}=0$. 
We also expand the auxiliary fields $\{ \varphi , \lambda_0 , \lambda^i \}$ and the matter field $\chi$ as
\begin{equation}
    \begin{aligned}
        \label{eq:15}
        \varphi & = \phi(t) + \delta\varphi(t, \bm{x})
        \,, &
        \lambda_0 & = \bar\lambda(t) + \delta\lambda_0(t, \bm{x})
        \,, \\
        \chi & = \bar\chi(t) + \delta\chi(t, \bm{x})
        \,, &
        \lambda^i & = \frac{1}{a^2} \delta^{ij} \partial_j \delta\lambda_s(t, \bm{x})
        \,.
    \end{aligned}
\end{equation}

The quadratic action for the tensor perturbation remains the same as in standard GR,
\begin{equation}
    S^{(2)}_T = \frac{\Mpl^2}{8} \int \md \tau\, \md^3k\, a^2 \left( \vert h_{ij}^\prime \vert^2 - k^2 \vert h_{ij} \vert^2 \right)\,.
    \label{eq:Tensor_action}
\end{equation}
The scalar quadratic action differs from that of a canonical matter field because of the nontrivial constraint structure.
It is convenient to introduce the comoving curvature perturbation,
\begin{equation}
    \mathcal{R}
    \equiv
    \zeta-\frac{\mathcal{H}}{\bar\chi^\prime}\delta\chi\,.
    \label{eq:Definition_Curvature_Linear}
\end{equation}
At linear order, this combination is invariant under
$x^\mu\to x^\mu+\xi^\mu$. This kinematic invariance does not, however,
restore the broken time-slicing symmetry of the action.
We define the background logarithmic derivatives
\begin{align}
    \eta & \equiv\frac{\dot \epsilon}{\epsilon H} \,, &
    \beta_\alpha & \equiv\frac{\dot \alpha}{\alpha H}\,.
\end{align}
For $k\neq0$, solving the linear constraint equations gives
\begin{align}
    & \delta\lambda_0=\delta\lambda_s=\delta\varphi=0\,, \\
    & \beta
    =\frac{3\alpha\mathcal{H}^2}{2k^2\bar\chi^\prime}\delta\chi\,,
    \label{eq:beta_linear}\\
    & \delta N
    =-\frac{2k^2}{\alpha\mathcal{H}^2}\zeta
    -\frac{\mathcal{H}^3\alpha(6+\beta_\alpha-2\epsilon)
    +a^2\bar\chi^\prime\Mpl^{-2}U_{,\chi}}
    {\mathcal{H}^2\alpha\bar\chi^\prime}\delta\chi
    +\frac{\delta\chi^\prime}{\bar\chi^\prime}\,,
    \label{eq:delta_N_linear}\\
    & \delta\chi
    =-\frac{\bar\chi^\prime}{\mathcal{H}}
    \frac{4k^2\mathcal{R}
    +2\alpha\mathcal{H}\mathcal{R}^\prime}
    {4k^2+\mathcal{H}^2\alpha(6+\alpha-2\epsilon)}\,.
    \label{eq:delta_chi_sol}
\end{align}
Substituting these solutions back into the action yields the quadratic
action for $\mathcal{R}$,
\begin{equation}
\label{eq:38}
    S_{\mathcal{R}^2}^{(2)}
    =\frac{\Mpl^2}{2}\int \md\tau\,\md^3k\,z^2
    \left(
    |\mathcal{R}_k^\prime|^2
    -c_{\mathcal R}^2k^2|\mathcal{R}_k|^2
    \right)\,,
\end{equation}
where
\begin{align}
    z^2 & = 2 a^2 \alpha \frac{2 k^2 + 3  \alpha \mathcal{H}^2 }{ 4 k^2 + 6  \alpha  \mathcal{H}^2 +  \alpha \mathcal{H}^2 ( \alpha - 2 \epsilon)  } \label{eq:Definition_zsq}\,, \\
    c_\mathcal{R}^2 & = \frac{8  k^4 + k^2 \mathcal{H}^2 B_1 + \mathcal{H}^4 B_2}{8  k^4 +  k^2 \mathcal{H}^2 A_1 + \mathcal{H}^4 A_2} \label{eq:definition_cR}\,,
\end{align}
with
\begin{align}
    B_1 & = 2  \Big[  \alpha \left(  - 4 \beta_\alpha + 2 \epsilon + \alpha \right) + 4 \epsilon \left(6 + \beta_\alpha -2 \epsilon + \eta  \right) \Big]\,, \\
    B_2 & = \alpha \Big[ - ( \alpha - 2\epsilon)^2 + 36  \epsilon + 6 ( \alpha (-1  - \beta_\alpha + \epsilon)  + 2 \epsilon ( 1 - \epsilon + \eta ) )    \Big]\,, \\
    A_1 & = 2 \alpha ( 12  + \alpha - 2 \epsilon )\,, \\
    A_2 & = 3 \alpha^2 ( 6  + \alpha - 2 \epsilon )\,.
\end{align}
For later purposes, it is convenient to introduce the dimensionless parameters $\tilde{\mathcal{H}} = -  \tau_B \mathcal{H} $ and $\kappa = - k \tau_B$.
Then the linear mode equation depends only on $\kappa$, $\alpha_0$, and $d$.

\subsection{Stability}

\subsubsection{\texorpdfstring{Evolution of $z$}{Evolution of z}}

For $\epsilon \leq 0$ and $\alpha > 0$, both the numerator and denominator
of $z^2$ are positive. Hence $z^2 > 0$ throughout Genesis, and the scalar
sector is ghost-free. At fixed nonzero $\kappa$ and $0 < d < 1$, one has
$\alpha \to \infty$ but $\alpha \tilde{\mathcal{H}} \to 0$ in the far past.
Therefore,
\begin{equation}
    z^2(x\to\infty)\simeq a^2\alpha \,,
\end{equation}
and $z^2$ diverges as $x\to\infty$.
This approximation also holds throughout Genesis for modes satisfying
\begin{equation}
    \kappa\gg
    \max(\alpha,\sqrt{\alpha},\sqrt{-\alpha\epsilon})\,
    \tilde{\mathcal{H}} \,.
\end{equation}
The strong-coupling issue and its absence are discussed in section~\ref{sec:Cubic_Interaction}.

\begin{figure}
    \centering
    \includegraphics[width=0.45\linewidth]{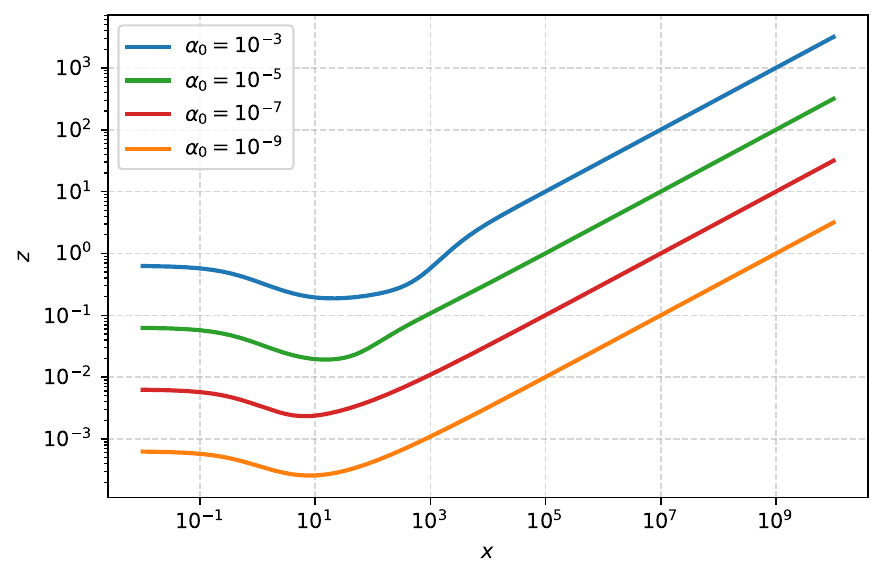}
    \includegraphics[width=0.45\linewidth]{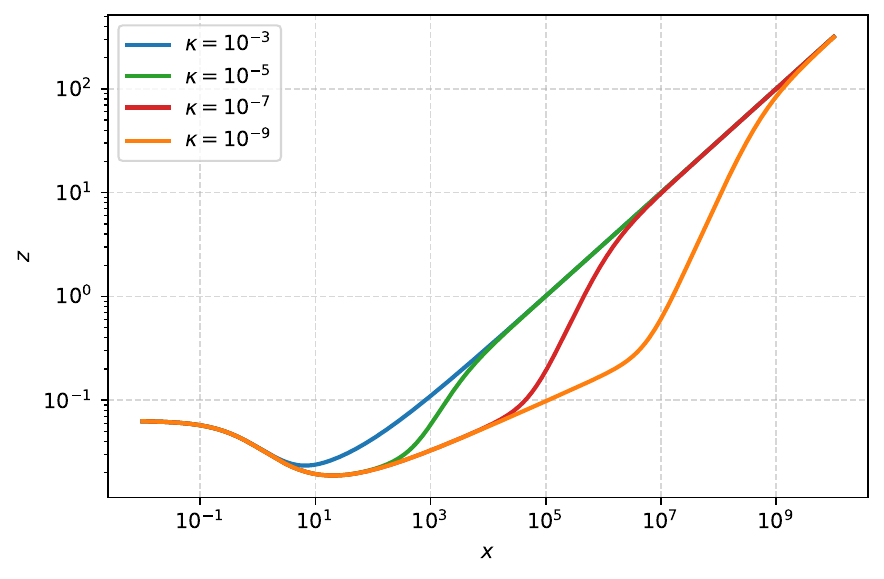} \newline
    \includegraphics[width=0.45\linewidth]{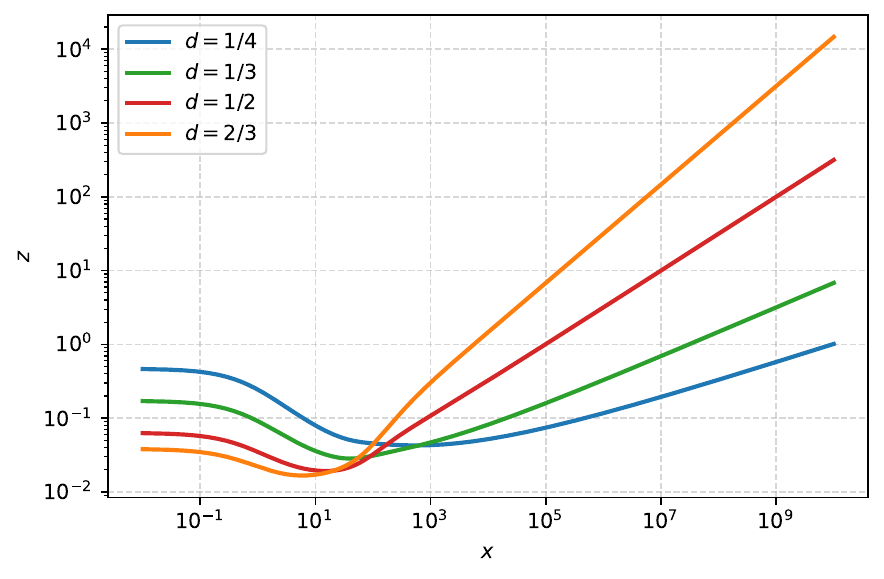}
    \caption{Evolution of $z(x)$ with respect to $x$. Upper left: varying $\alpha_0$ at $\kappa=10^{-4}$ and $d=1/2$; upper right: varying $\kappa$ at $d=1/2$ and $\alpha_0=10^{-5}$; bottom: varying $d$ at $\alpha_0=10^{-5}$ and $\kappa=10^{-4}$.}
    \label{fig:Evolution_z}
\end{figure}

During the early Genesis phase, $z^2$ decreases, leading to the later
superhorizon growth of $\mathcal{R}$. Near the end of Genesis, $z$ increases
with time, so $\mathcal{R}$ is suppressed, but it does not necessarily become
constant. Figure~\ref{fig:Evolution_z} exhibits this behavior.

\subsubsection{\texorpdfstring{Infrared region with $c_\mathcal{R}^2<0$}{Infrared region with negative squared sound speed}}

The squared sound speed depends on momentum and can become negative over a finite
range of $\kappa$. Since the denominator of equation~\eqref{eq:definition_cR}
is positive, the boundaries of this region are determined by the zeros
of the numerator,
\begin{equation}
\kappa_\pm^2
=
\frac{\tilde{\mathcal{H}}^{\,2}}{16}
\left[
-B_1\pm\sqrt{B_1^2-32B_2}
\right]\,,
\qquad
\max(0,\kappa_-^2)<\kappa^2<\kappa_+^2\,.
\end{equation}
This region exists when $B_1^2-32B_2>0$ and $\kappa_+^2>0$.

A negative $c_\mathcal{R}^2$ does not by itself imply that the canonical
mode is non-oscillatory, since its evolution is governed by
\begin{equation}
\omega^2(\kappa,x)
=
c_\mathcal{R}^2\kappa^2-\frac{z_{,xx}}{z}\,.
\end{equation}
Therefore, the regions with $c_\mathcal{R}^2<0$ and $\omega^2<0$ do not
generally coincide.

\begin{figure}
    \centering
    \includegraphics[width=0.45\linewidth]{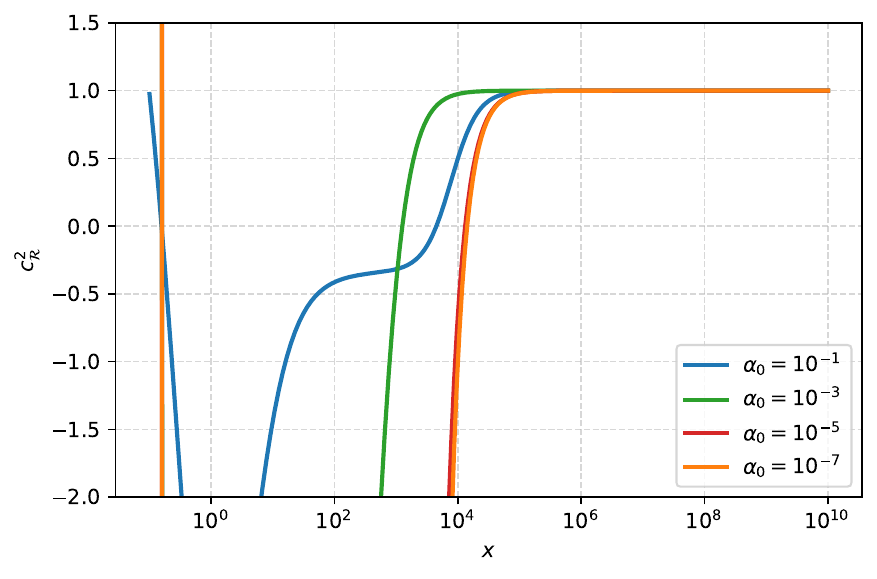}
    \includegraphics[width=0.45\linewidth]{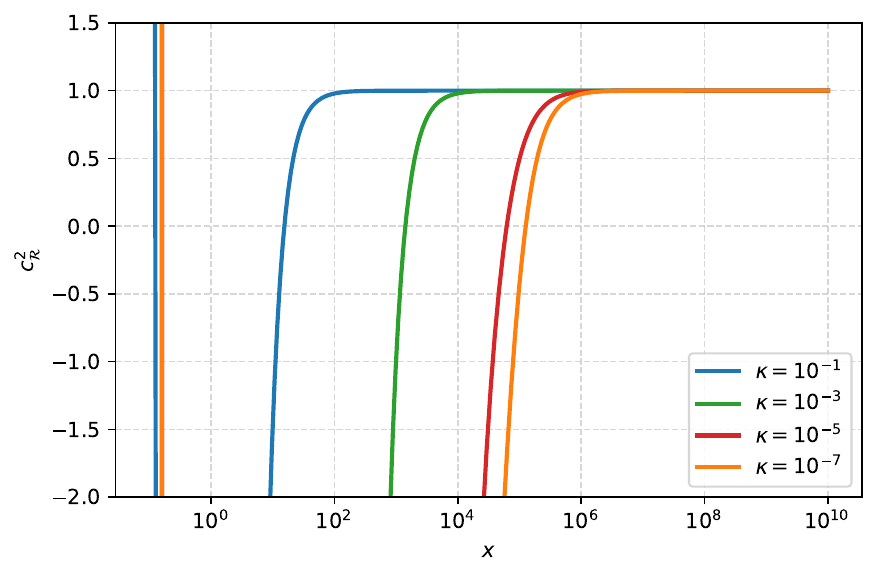}
    \includegraphics[width=0.45\linewidth]{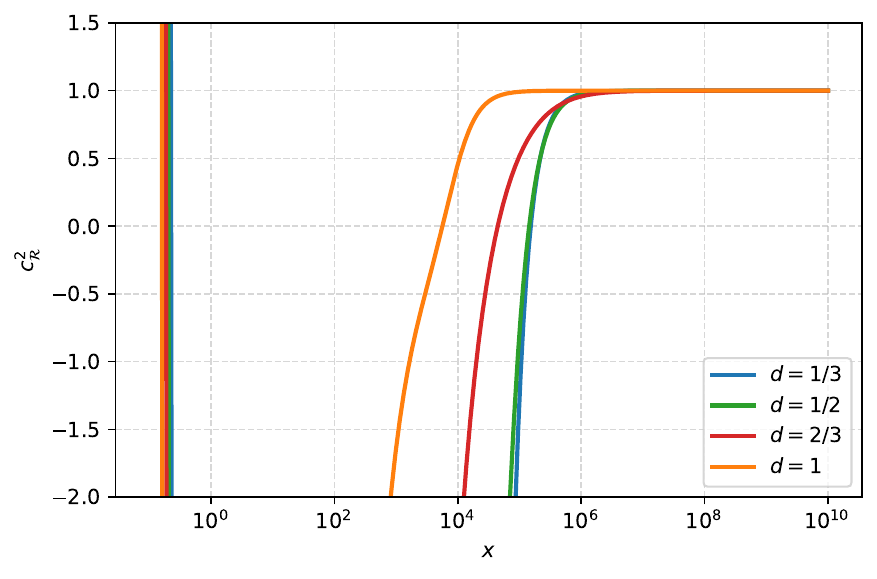}
    \caption{Evolution of $c_\mathcal{R}^2(x)$. Upper left: varying $\alpha_0$ at $d=1$ and $\kappa=10^{-4}$; upper right: varying $\kappa$ at $d=1$ and $\alpha_0=10^{-5}$; bottom: varying $d$ at $\kappa=10^{-5}$ and $\alpha_0=10^{-3}$. At finite momentum, $c_\mathcal{R}^2$ may be negative or exceed unity.}
    \label{fig:Evolution_c_R}
\end{figure}

In the far past, for $\alpha\gg\max(1,|\epsilon|,|\beta_\alpha|)$, the upper boundary behaves as
\begin{equation}
\kappa_+^2
\simeq
\frac{\alpha_0(1+d)^2}{2}(1+x)^{-2}\,.
\end{equation}
The expansion requires $\alpha_0(1+x)^d\gg1$, equivalently $\alpha\gg|\epsilon|$, which is satisfied in the stated regime.
Hence the negative-$c_\mathcal{R}^2$ region shrinks toward zero momentum.
Thus, every fixed nonzero mode eventually lies outside this region in the
deep Genesis phase. For $\kappa^2\gg\alpha(6+\alpha-2\epsilon)\tilde{\mathcal{H}}^{\,2}$, the large-momentum expansion gives
\begin{equation}
c_\mathcal{R}^2\kappa^2
=
\kappa^2
+
\frac{(B_1-A_1)\tilde{\mathcal{H}}^{\,2}}{8}
+
\mathcal{O}(\kappa^{-2})\,,
\label{eq:Sound_speed_expansion}
\end{equation}
and hence $c_\mathcal{R}^2\to1$ in the ultraviolet. The negative region is
therefore an infrared effect rather than an ultraviolet gradient
instability.
Figure~\ref{fig:Evolution_c_R} shows the evolution of $c_\mathcal{R}^2$.
Although $c_\mathcal{R}^2$ can exceed unity at finite momentum, this can arise from a mass-like correction and does not by itself
imply superluminal propagation. Furthermore, even superluminal propagation would not violate causality in the sense of past and future defined by the VCDM preferred frame. 

\subsection{Evolution of the curvature perturbation and power spectrum}

For the canonically normalized variable $u=\Mpl z\mathcal{R}$, the mode equation is
\begin{equation}
    \frac{\md^2 u}{\md x^2}
    +
    \left(
    c_\mathcal{R}^2\kappa^2
    -
    \frac{z_{,xx}}{z}
    \right)u
    =0\,.
\end{equation}
For the numerical spectra, we define $\widehat u_k=u_k/\sqrt {-\tau_B}$.
No closed-form solution is available, so we use analytic approximations
to set initial conditions and then solve the full equation numerically.

We impose initial conditions sufficiently far in the past so that
$|c_\mathcal{R}^2-1|\ll1$, $|z_{,xx}/z|\ll\kappa^2$, and so that the evolution is
adiabatic. In this limit, we have that $z_{,xx} \cdot z^{-1} \simeq 0$ and $c_\mathcal{R}^2\simeq1$.
Therefore, the canonical mode approaches a Minkowski plane wave, that is
\begin{equation}
    u \vert_{x \rightarrow \infty}
    \simeq
    \sqrt{-\tau_B}\,
    \frac{e^{i\kappa x}}{\sqrt{2\kappa}}\,,
    \qquad
    \left.
    \frac{\md u}{\md x}
    \right|_{x\to\infty}
    \simeq
    i\sqrt{-\tau_B}\sqrt{\frac{\kappa}{2}}\,
    e^{i\kappa x}\,.
    \label{eq:Initial_conditions}
\end{equation}
Equation~\eqref{eq:Initial_conditions} specifies the asymptotic
Minkowski vacuum. In the numerical calculation, the evolution begins
at a finite time $x_i$ inside the adiabatic oscillatory regime.
We therefore use
\begin{align}
    \widehat u_k(x_i) &
        =
        \frac{1}{\sqrt{2\Omega_i}}\,,
        \\
    \widehat u_{k,x}(x_i) &
        =
        \left[
        i\Omega_i
        -
        \frac{\Omega_{,x}(x_i)}{2\Omega_i}
        \right]
        \widehat u_k(x_i)\,,
\end{align}
where we defined
\begin{subequations}
    \begin{align}
        \Omega_i & \equiv \Omega(x_i)\,, \\
        \Omega(x) & \equiv \sqrt{\omega^2(x)} \,.
    \end{align}
\end{subequations}
These initial data reduce to the plane-wave form in
equation~\eqref{eq:Initial_conditions} as $x_i\to\infty$.
In the parameter range considered here, $0<d<1$, the same asymptotic regime
gives $z^2\simeq a^2\alpha$, and hence
$|\mathcal{R}|=|u/(\Mpl z)|\sim x^{-d}$.

At intermediate times the effective mass term cannot be neglected.
Near the non-oscillatory transition, we use the canonical turning point
$\omega^2 = 0$, which generally differs from Hubble radius crossing
$\kappa=\tilde{\mathcal{H}}$. At small $\alpha_0$ and with both $T$ and
$\kappa$ fixed, the exact coefficients give
\begin{equation}
    \omega^2
    =
    \kappa^2
    -
    \frac{
    2+3 \, (1+d)^2 \, T^{-d} \, (1-T^{-d})
    }{T^2}
    +
    \mathcal{O}(\alpha_0)\,.
\end{equation}

The leading Hankel form further requires
\begin{align}\
    \alpha & \ll 1 \,, &
    \alpha\epsilon^2 & \ll 1 \,, &
    x & \gg 1 \,, &
    3(1+d)^2T^{-d}(1-T^{-d}) & \ll2 \, .
\end{align}
Under these conditions, we have
\begin{equation}
    \omega^2
    =
    c_\mathcal{R}^2\kappa^2
    -
    \frac{z_{,xx}}{z}
    \simeq
    \kappa^2-\frac{2}{x^2}\,.
\end{equation}
The corresponding Hankel solution gives the leading scale-invariant
spectrum. Under the same assumptions, with
$a\simeq1$ and $(1+d)T^{-d}/d\ll1$, the growing curvature mode scales as
$|\mathcal{R}|\sim x^{-1-d}$.
Toward the end of the Genesis phase, the modes approach a finite
amplitude, although $\mathcal{R}$ is not exactly conserved.

\begin{figure}
    \centering
    \includegraphics[width=0.45\linewidth]{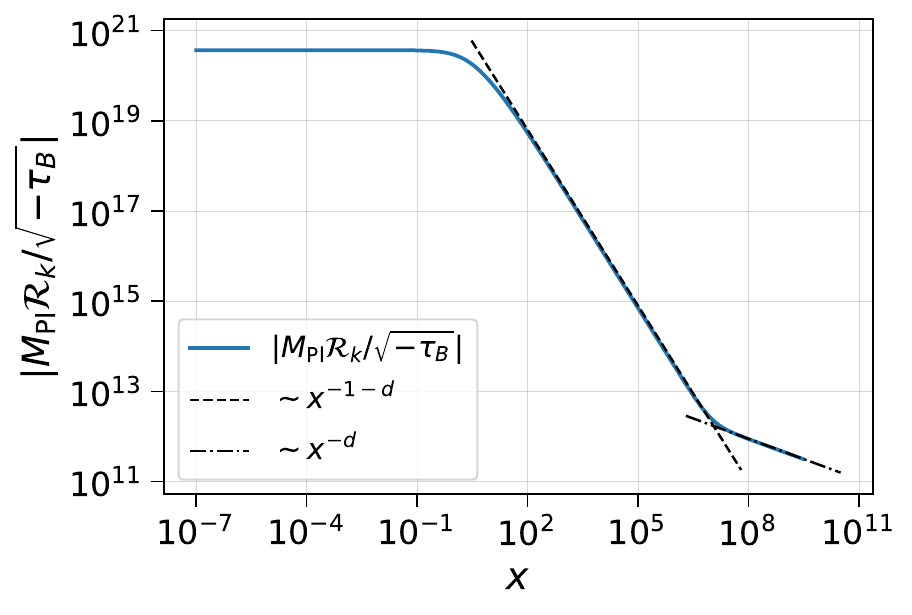}
    \includegraphics[width=0.45\linewidth]{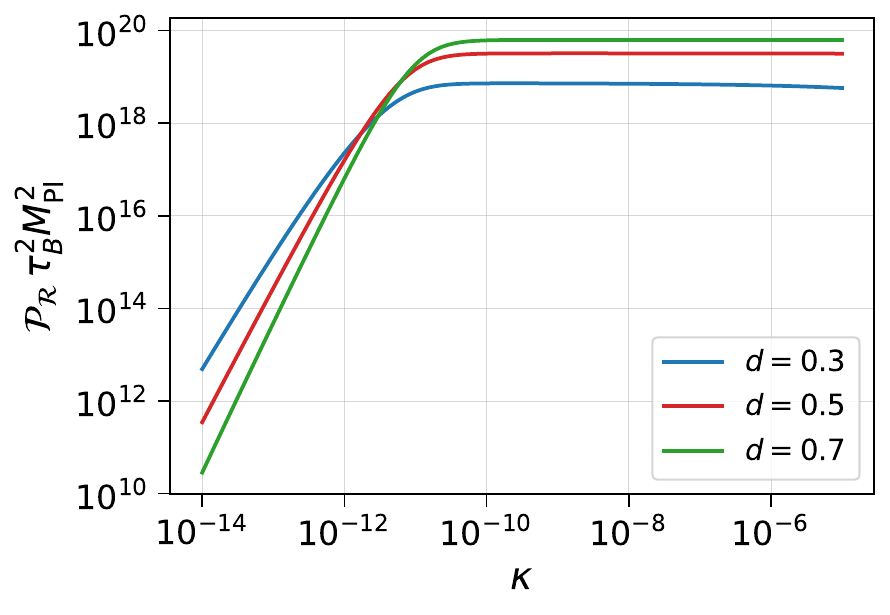}
    \caption{
    Left: evolution of $\mathcal{R}$ for $d=0.3$, $\alpha_0=10^{-22}$,
    and $\kappa=10^{-7}$. The black curves show the $x^{-1-d}$ and
    $x^{-d}$ asymptotic scalings.
    Right: scalar power spectra for $\alpha_0=10^{-22}$ and
    $d=0.3$, $0.5$ and $0.7$.
    }
    \label{fig:Evolution_Power_R}
\end{figure}

The left panel of figure~\ref{fig:Evolution_Power_R} shows the numerical
evolution of $|\mathcal{R}|$ for $d=0.3$, $\alpha_0=10^{-22}$, and
$\kappa=10^{-7}$.
One can visually observe the close agreement between the asymptotic
behavior with its analytical estimates.
We evaluate $|\mathcal{R}|$ at $x_f=10^{-7}$, close to the end of the
Genesis phase at $x=0$. The power spectrum is
\begin{equation}
    \mathcal{P}_\mathcal{R}
    =
    \frac{\kappa^3}{2\pi^2\tau_B^2\Mpl^2}
    \left|
        \frac{\widehat u_k(x_f)}{z(x_f)}
    \right|^2\,.
    \label{eq:PR_numerical_endpoint}
\end{equation}
The spectrum is evaluated at $x_f$ after including the superhorizon
evolution. Matching to a later phase requires both $\mathcal{R}$ and its
conjugate momentum. For $d=0.3$, $\alpha_0=10^{-23}$, and
$\kappa=10^{-7}$, the power at $x_f$ differs from its $x=0$ value by
only $2.27\times10^{-8}$, although
$(\md\ln|\mathcal{R}|/\md\ln a)_0=0.08727$.
Initial data are imposed only in the adiabatic oscillatory regime, where
\begin{align}
    \omega^2 & > 0 \, , &
    |c_\mathcal{R}^2-1| & \ll 1 \,, &
    |z_{,xx}/z|/\kappa^2 & \ll 1 \,, &
    |\omega_{,x}|/\omega^2 & \ll 1 \,.
\end{align}

The right panel of figure~\ref{fig:Evolution_Power_R} shows the scalar
power spectrum of several values of $d$. Their overall amplitude can be
adjusted with $\tau_B$. The infrared spectrum is strongly blue, while an
intermediate range is red.

We scan this intermediate regime on an $80^3$ grid in
$(\kappa,\alpha_0,d)$:
\begin{equation}
    10^{-30}\leq\alpha_0\leq10^{-21}\,,
    \qquad
    0.1\leq d\leq0.4\,,
    \qquad
    10^{-9}\leq\kappa\leq10^{-4}\,.
\end{equation}
Appendix~\ref{app:Phenomenological_fit} gives a three-parameter
phenomenological fit to the power spectrum.

The scalar spectral index and its running can be obtained directly from the fit,
\begin{equation}
    n_s^{\text{fit}}-1
    =
    \frac{\md\ln\mathcal{P}_\mathcal{R}^{\text{fit}}}
         {\md\ln\kappa}\,,
    \qquad
    \alpha_s^{\text{fit}}
    =
    \frac{\md n_s^{\text{fit}}}
         {\md\ln\kappa}\,.
\end{equation}
Since $\kappa=-k \, \tau_B$, this is equivalent to differentiating with
respect to $\ln k$ at fixed $\tau_B$.

\begin{figure}
    \centering
    \includegraphics[width=0.45\linewidth]{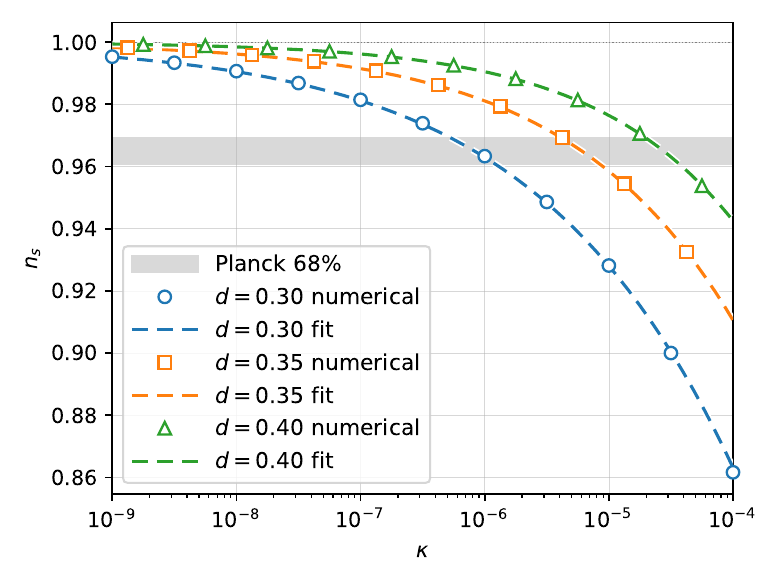}
    \includegraphics[width=0.45\linewidth]{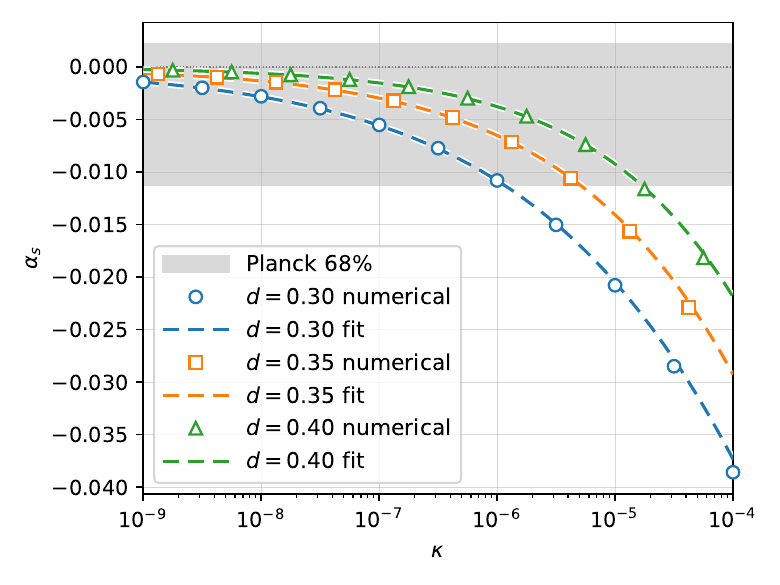}
    \caption{
    Spectral index $n_s$ (left) and running $\alpha_s$ (right) for
    $\alpha_0=10^{-23}$. Dots are derivatives of the numerical
    spectrum, without smoothing; Dashed curves are obtained from the
    phenomenological fit. The running panel uses $d = 0.30$, $0.35$ and $0.40$.
    Gray bands show the quoted Planck $68\%$ intervals.
    }
    \label{fig:Spectral_Index_and_running}
\end{figure}

Figure~\ref{fig:Spectral_Index_and_running} compares the fit derivatives
with derivatives obtained directly from the numerical spectrum.
Here, we have$\alpha_s=\md n_s/\md\ln k$.
Across the displayed range, the fit differs from the numerical
derivatives by at most $1.5\times10^{-3}$ in $n_s$ and
$1.3\times10^{-3}$ in $\alpha_s$.
For reference, the quoted Planck constraints~\cite{Planck:2018jri} are
\begin{equation}
    n_s = 0.9649 \pm 0.0042\,,
    \qquad
    \alpha_s = -0.0045 \pm 0.0067\,.
\end{equation}
The quoted Planck values are separate $68\%$ marginal constraints:
$n_s$ is obtained without running, whereas the $\alpha_s$ constraint
allows running. The mapping between $\kappa$ and the observed pivot
depends on $\tau_B$ and the post-Genesis evolution, so these values are
shown only as a reference.

\subsection{Tensor modes}

\begin{figure}
    \centering
    \includegraphics[width=0.45\linewidth]{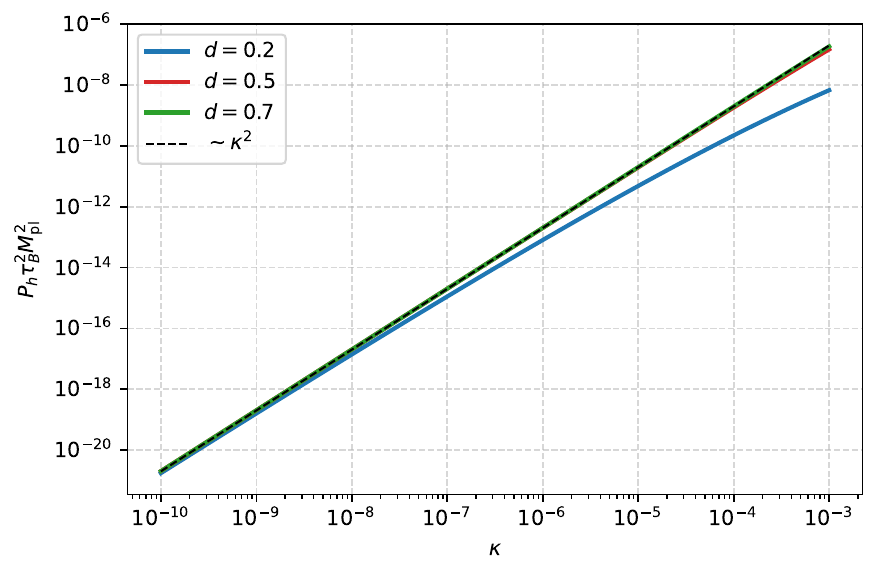}
    \caption{Tensor power spectrum for $d=0.2$, $0.5$, and $0.7$. The black dashed line shows the $\kappa^2$ scaling.}
    \label{fig:P_h}
\end{figure}

Because of the constraint structure of VCDM, tensor modes are unaffected, and their quadratic action is equation~\eqref{eq:Tensor_action}, the same as in GR. For the canonically normalized mode $u^\lambda=a\Mpl h^\lambda/2$,
\begin{equation}
    \frac{\md^2 u^\lambda}{\md x^2} + \left( \kappa^2 - (1+d)^2 \frac{1 + (1+x)^d}{(1+x)^{2+2d}} \right)  u^\lambda =0\,.
\end{equation} For either polarization, define $\widehat u_T=u^\lambda/\sqrt{-\tau_B}$ and normalize $e_{ij}^{\lambda}e_{ij}^{\sigma *}=\delta_{\lambda\sigma}$. The total power sums the two equal polarizations.
The power spectrum for the gravitational waves is given by
\begin{equation}
    \mathcal{P}_h = \frac{4 \kappa^3 \big \vert \frac{\widehat u_T(0)}{a(0)} \big\vert^2}{ \pi^2 \tau_B^2 \Mpl^2}\,.
\end{equation}
Figure~\ref{fig:P_h} shows the numerical spectrum for several values of $d$.
With this normalization, an independent numerical evolution reproduces figure~\ref{fig:P_h}.

At leading order, the tensor-to-scalar ratio is $r=\mathcal{P}_h/\mathcal{P}_\mathcal{R}\propto\alpha_0\ll1$, and we therefore observe that $r$ is strongly suppressed.

\subsection{Geodesic completeness}

To discuss geodesic completeness, we consider the tensor and scalar propagations separately.

The tensor action \eqref{eq:Tensor_action} can be written as
\begin{equation}
    S^{(2)}_T = - \frac{\Mpl^2}{8} \int \md \tau\, \md^3x\, a^2 \eta^{\mu\nu}  \partial_\mu h_{ij} \partial_\nu h_{ij}\,,
\end{equation}
where $\eta^{\mu\nu} $ is the Minkowski metric.
Therefore, as expected, the effective metric for the tensor perturbation is past geodesically complete. Indeed, $a\to1$ as $\tau\to-\infty$, so both integrals $\int a^2\md\tau$ and $\int a^2\md\tau/\sqrt{a^2+P^2}$ diverge, where $P$ is the conserved comoving momentum.

The scalar sector is subtler because its kinetic coefficient and sound speed depend on momentum. A variable-dependent effective metric cannot by itself define spacetime completeness, and VCDM preserves only foliation-preserving diffeomorphisms. We therefore also use the quadratic action in a regular variable $\delta\chi$, which is derived in appendix~\ref{app:Quadratic_action_delta_chi}.

The quadratic actions for $\mathcal{R}$ and $\delta\chi$ are
\begin{equation}
    S_{\mathcal{R}^2}^{(2)} = \frac{\Mpl^2}{2}\int \md \tau\, \md^3k  \, z^2 \left(
        \vert \mathcal{R}_k^\prime \vert^{2}
        - c_\mathcal{R}^2 k^2 \vert \mathcal{R}_k\vert^2
    \right) \,,
\end{equation}
and
\begin{equation}
    S_{\delta \chi^2}^{(2)} = \frac{1}{2} \int \md^3k \md \tau\, a^2 K(k,\tau) \left( \vert \delta \chi^{\prime} \vert^{2} - c_{\delta \chi}^2(k,\tau)  k^2 \vert \delta \chi \vert^2 \right)\,.
\end{equation}
For $k \ne 0$, $z^2 > 0$ throughout Genesis, while the kinetic coefficient $K$ of the $\delta\chi$ action has a pole at $k_\star^2=\alpha\mathcal{H}^2/2$. This pole reflects the velocity reduction, not a singularity of the canonical phase-space evolution. Because $\delta\chi$ mixes the curvature coordinate and momentum, $K<0$ does not imply a UV ghost instability (see \cite{Gumrukcuoglu:2016jbh}), and the region in the Fourier space where $K<0$ needs not coincide with the region where $c_\mathcal{R}^2<0$ or with the canonical non-oscillatory regime. Because the exact coefficients depend on momentum, the scalar sector cannot generally be represented by a single local effective metric. In the far past, an approximately massless $\delta\chi$ regime requires both $k\gg k_\star$ and $k (-\tau- \tau_B )\gg\sqrt{2(1+\chi_1^2/\Mpl^2)}$. For $\mathcal{R}$, the requirement is $k^2\gg\alpha(6+\alpha-2\epsilon)\mathcal{H}^2$. In these domains,
\begin{align}
    & z^2 \simeq a^2 \alpha\,, \qquad & c_\mathcal{R}^2 k^2 \simeq k^2 + M_\mathcal{R}^2\,, \\
    & K \simeq 1\,, \qquad & c_{\delta\chi}^2 k^2 \simeq k^2\,.
\end{align}
Here $M_\mathcal{R}^2=(B_1-A_1)\mathcal{H}^2/8$.
The curvature mode is approximately massless only when
$|M_\mathcal{R}^2|\ll k^2$.
The corresponding asymptotic quadratic actions are
\begin{align}
    S^{(2)}_{\mathcal{R}^2} \simeq & - \frac{\Mpl^2}{2} \int \md^4x\,a^2 \alpha \left(  \eta^{\mu\nu} \partial_\mu \mathcal{R} \partial_\nu \mathcal{R} + M_\mathcal{R}^2 \mathcal{R}^2 \right)\,, \\
    S^{(2)}_{\delta\chi^2} \simeq & - \frac{1}{2} \int \md^4x\, a^2 \eta^{\mu\nu} \partial_\mu \delta \chi \partial_\nu \delta \chi\,.
\end{align}
Within these approximations, the matter kinetic metric approaches $a^2\eta_{\mu\nu}$. For $0<d<1$, the curvature kinetic metric has conformal factor $a^2\alpha$, and $\int a^2\alpha\,\md\tau$ diverges in the past. Every fixed nonzero mode eventually enters this asymptotic regime, but the exact momentum-dependent scalar dynamics is not described by a single local metric at all wavelengths.

\section{Cubic interactions}
\label{sec:Cubic_Interaction}

Strong coupling is a common concern in Genesis models (see, e.g.,~\cite{Ageeva:2018lko,Ageeva:2020gti,Ageeva:2020buc,Ageeva:2021yik,Cai:2022ori,Akama:2022usl,GilChoi:2025hbs}).
We therefore examine the cubic interactions.
The linear definition \eqref{eq:Definition_Curvature_Linear} must be completed at second order. The temporal terms follow~\cite{Lyth:2005du,Ganz:2024ihb}, while the gradient term is obtained by transforming to uniform-field slices and restoring the spatial gauge. For scalar perturbations only, we obtain
\begin{align}
    \mathcal{R}_{1} &
        = \zeta_1 - \frac{\mathcal{H}}{\bar \chi^\prime} \delta \chi_1\,, \\
    \mathcal{R}_{2} &
        = \zeta_2 -  \frac{\mathcal{H}}{\bar \chi^\prime}  \delta \chi_2 + \frac{2 \mathcal{H}}{\bar \chi^{\prime 2}} \delta \chi_1 \delta \chi_1^\prime - 2 \frac{\delta \chi_1}{\bar \chi^\prime } \zeta_1^\prime  - \frac{\beta_\alpha}{2} \left( \frac{\mathcal{H} \delta \chi_1}{\bar \chi^\prime}\right)^2 +\Delta\mathcal{R}_2^{(\nabla)}\,.
\end{align}
With the convention $\delta\chi=\delta\chi_1+\delta\chi_2/2$, and similarly for $\zeta$ and $\mathcal{R}$, the additional nonzero-momentum gradient term is
\begin{align}
    \Delta\mathcal{R}_2^{(\nabla)} & =\frac12\left(\delta^{ij}-\frac{\partial^i\partial^j}{\Delta}\right)X_{ij} \, , \\
    \theta_1 & =-\frac{\delta\chi_1}{\bar\chi^\prime} \,,\\
    X_{ij} & =\partial_i\beta\,\partial_j\theta_1+\partial_j\beta\,\partial_i\theta_1-\partial_i\theta_1\,\partial_j\theta_1.
\end{align}
Here $\mathcal{H}/\bar\chi^{\prime2}=(1+d)(-\tau_B)T^{1-d}/\chi_1^2$ diverges only for $0<d<1$, while $(\bar\chi^\prime)^{-2}\propto T^2$ grows for all $d>0$. These growing coefficients make the nonlinear conversion nonuniform at fixed momentum, but they neither determine the full second-order solution nor establish physical strong coupling. The linear curvature mode still vanishes in the far past.
Using $\mathcal{R}=\mathcal{R}_1+\tfrac12\mathcal{R}_2$ to diagnose strong coupling can therefore be misleading.

Bouncing and Genesis models often avoid this problem by working in the flat gauge, but that gauge is unavailable in VCDM because time diffeomorphism invariance is broken. We instead analyze the regular variable $\delta\chi$ in a chosen spatial gauge. Appendix~\ref{app:Quadratic_action_delta_chi} gives its quadratic action and constraint structure. At fixed nonzero momenta, $a\sqrt K\to1$ in the far past, so $\delta\chi$ has a regular asymptotic normalization.

\subsection{Absence of strong coupling below $\Mpl$}

Taking $\delta\chi$ as the fundamental variable, the linear constraints are
\begin{align}
    \delta N & =\frac{\delta \chi'}{\bar\chi'}+ \frac{a^2U_{,\chi}}{\bar\chi'^2} \delta \chi +\Dsp\zeta\,, \\
    \beta & =-\frac{3\bar\chi'}{2\Mpl^2}\Delta^{-1}\delta \chi \, , \\
    \zeta & = \frac{(3-\Dsp)\frac{\delta\chi'}{\bar\chi'}-(3+\Dsp) \frac{a^2U_{,\chi}}{\bar\chi'^2} \delta \chi }{\Dsp(1+\Dsp)}\,,
\end{align}
where
\begin{equation}
    \Dsp\equiv\frac{2\Mpl^2}{\bar\chi'^2}\Delta\,.
\end{equation}
At fixed nonzero momentum, the deep-Genesis limit $x\to\infty$ gives
\begin{equation}
    |\Dsp|=2\Mpl^2k^2 \tau_B^2 T^2/\chi_1^2\gg1 \,,
\end{equation}
and hence
\begin{align}
    \delta N & \sim \mathcal{O}( (\bar \chi^\prime\Dsp)^{-1}) \,,
    & \beta  & \sim \mathcal{O}( (\bar \chi^\prime\Dsp)^{-1} ) \,, &
    \zeta & \sim \mathcal{O}( (\bar \chi^\prime\Dsp)^{-1}) \,.
\end{align}
Thus, at fixed nonzero momentum, $\delta N$, $\beta$, and $\zeta$
vanish in the far past. Appendix~\ref{app:Cubic_action_Power_counting}
shows that the pure-scalar cubic interactions are suppressed as
$x\to\infty$.

Keeping only the leading terms, we obtain the cubic scalar interaction
\begin{align}
    S^{(3)}_{\delta\chi^3} \simeq & \int \md^4x\, a^2 \Bigg\{ \frac12(3\zeta-\delta N) \delta\chi'^2
    -\delta\chi'
    \partial_i\beta\,\partial_i \delta\chi
    -\frac{\delta N+\zeta}{2}(\partial_k \delta\chi)^2
    \Bigg\} \nonumber \\
    \simeq & \frac{\chi_1}{4 \Mpl^2 (-\tau_B) } \int \md^4x\, \frac{1}{1+x} \Big[ 7(\delta \chi'_c)^2\Delta^{-1}\delta\chi'_c
    -6\delta\chi'_c\,\partial_i\Delta^{-1}\delta\chi_c\,\partial_i\delta\chi_c
    +3(\Delta^{-1}\delta\chi'_c)(\partial_i\delta\chi_c )^2 \Big]\,.
\end{align}
Here, $\delta\chi_c=a\sqrt K\,\delta\chi$ is canonically normalized, with $\sqrt K$ acting mode by mode on the $K>0$ branch. The approximation $\delta\chi_c\simeq a\delta\chi$ holds only asymptotically at fixed nonzero momentum and is not an exact local field redefinition at finite time. Since the cubic scalar interaction vanishes when $x\to\infty$, no cubic strong-coupling scale collapses to zero in this limit. This does not remove the ordinary matter--graviton interaction or establish finite-time, all-order control.

The mixed scalar--tensor interactions are analyzed in appendix~\ref{app:Cubic_action_Power_counting}, while the pure tensor cubic action is the GR one, which reads
\begin{equation}
    S^{(3)}_{h^3} = \frac{\Mpl^2}{4} \int \md^4x\, a^2 \left( h_{ik} h_{j\ell} - \frac12 h_{ij} h_{k\ell} \right) \partial_k \partial_\ell h_{ij}\,.
\end{equation}
After canonical normalization, this interaction has the usual strength $E/\Mpl$ and is weak for $E\ll\Mpl$.
The scalar--scalar--tensor interaction in the deep Genesis limit scales as
\begin{equation}
    S^{(3)}_{\delta\chi^2 h} \simeq  \int \md^4x\,  \frac{1}{2} h^{ij} \partial_i \delta \chi_c \partial_j \delta \chi_c\,,
\end{equation}
which is the only asymptotically non-vanishing contribution within the two-scalar/one-tensor sector.
Finally, the tensor--tensor--scalar interactions vanish asymptotically, with the same $\mathcal{O}((\bar\chi^\prime\Dsp)^{-1})$ suppression as the pure scalar sector.

The leading cubic interactions therefore scale as given in table~\ref{tab:final_sector_table_new}.
\begin{table}[H]
    \centering
    \caption{Leading cubic interactions scalings.}
    \label{tab:final_sector_table_new}
    \begin{tabular}{cc}
        \toprule
        Sector & Coupling \\
        \midrule
        $\delta\chi_c^3$
        & $\mathcal O(x^{-1})$
        \\[2.5mm]
        $\delta\chi_c^2 h$
        & $\mathcal O(E/\Mpl)+\mathcal O(x^{-2})$
        \\[2.5mm]
        $\delta\chi_c h^2$
        & $\mathcal O(x^{-1})$
        \\[2.5mm]
        $h^3$
        & $\mathcal O(E/\Mpl)$ \\
        \bottomrule
    \end{tabular}
\end{table}
where $E = k/a$ is the physical momentum. Here $h_{c,ij}=a\Mpl h_{ij}/2$; the time scalings assume fixed parameters and generic nonzero momenta, excluding parametrically soft internal momenta.
Thus, all background-dependent cubic couplings vanish in the asymptotic
past, while the remaining matter--graviton and graviton self-interactions
are Planck suppressed. Therefore, the cubic strong-coupling scale is
parametrically
\begin{equation}
\Lambda_{\text{cubic}}\sim \Mpl\,.
\end{equation}
Thus, at cubic order there is no new cutoff besides the Planck mass $\Mpl$.

\section{Conclusion}
\label{sec:Conclusion}

We have constructed an asymptotically Minkowski Genesis cosmology in
VCDM coupled to a canonical scalar field. The modified gravitational
equations allow
\begin{align}
  \dot{H}/H^2 & > 0 \,, &
  \rho_\chi+p_\chi & =\dot{\bar\chi}^{\,2}>0\,,
\end{align}
so that the effective NEC-violating expansion does not require NEC
violation in the matter sector. The gravitational sector has only
two propagating tensor degrees of freedom. The background metric, which
also governs tensor propagation, is past geodesically complete.

The curvature perturbation has a positive kinetic coefficient throughout
the Genesis epoch, and its short-wavelength sound speed approaches unity,
$c_\mathcal{R}^2\to1$. At finite momentum, $c_\mathcal{R}^2$ can become
negative within an infrared band. This is not an ultraviolet gradient
instability and, in general, does not coincide with the non-oscillatory
regime of the canonical mode, which is determined by the sign of 
$\omega^2=c_\mathcal{R}^2\kappa^2-z_{,xx}/z$. For the parameter range
considered, the region with $c_\mathcal{R}^2<0$ shrinks toward zero
momentum in the asymptotic past, while every fixed nonzero mode satisfies
$\omega^2\to\kappa^2>0$ and therefore becomes oscillatory.

The numerical scalar spectrum exhibits a nearly scale-invariant,
slightly red-tilted regime over an intermediate range of scales. Under the controlled turning-point approximations of section~\ref{sec:Linear_perturbations}, the relevant modes obey
\begin{equation}
  \frac{\md^2u}{\md x^2}
  +\left(\kappa^2-\frac{2}{x^2}\right)u\simeq0\,,
\end{equation}
which accounts for the leading scale-invariant behavior. The curvature
perturbation then undergoes non-oscillatory growth, including a superhorizon stage for the modes studied, and approaches a finite amplitude at the end of Genesis. The tensor sector
has the GR quadratic action and produces a blue spectrum, with
\begin{equation}
  \mathcal{P}_h\propto\kappa^2\,,
  \qquad
  r\propto\alpha_0\ll1\,,
\end{equation}
for the parameter region studied.

The cubic analysis highlights the importance of a regular perturbation
variable. The nonlinear transformation to the curvature perturbation can
become singular in the deep Genesis limit, whereas the matter fluctuation
$\delta\chi$ has a regular asymptotic normalization. At fixed nonzero
momenta and away from the singular surface $\Dsp=-1$ of the matter-variable velocity reduction,
we find that the background-dependent cubic couplings vanish in the
asymptotic past. The surviving matter--graviton and graviton
self-interactions remain Planck suppressed. The resulting cubic power
counting gives $\Lambda_{\text{cubic}}\sim\Mpl$ in this limit, with no cubic
strong-coupling scale that collapses to zero toward the past. This
asymptotic result does not determine the full nonlinear effective field
theory cutoff.

A complete early-Universe scenario will require a systematic analysis of higher-order interactions, together with an explicit matching of both the background evolution and perturbations onto a subsequent inflationary phase and/or reheating epoch. Since the VCDM, as one of type-II minimally modified gravity theories, does not have any extra gravitational degrees of freedom beyond the ordinary tensor gravitons with $\Mpl$-suppressed interactions, it serves as a promising framework that potentially avoids strong coupling of higher-order interactions in early-universe cosmological scenarios. We leave these important questions for future work.

\acknowledgments

We would like to thank Ryo Namba for helpful comments and discussions.
The work of P.P. is supported by IBS under the project code IBS-R018-D3. A.G. is supported by the DFG under the Emmy-Noether program, project number 496592360.
P.M. is supported by grants from the Research Grants Council of Hong Kong (Project Nos. CUHK 14304622, 14307923, and 14307724), the start-up grant from The Chinese University of Hong Kong, and the Direct Grant for Research from the Research Committee of The Chinese University of Hong Kong. The work of S.M. was supported in part by Japan Society for the Promotion of Science Grants-in-Aid for Scientific Research No. 24K07017 and the World Premier International Research Center Initiative (WPI), MEXT, Japan.

\vspace{0.0cm}

\appendix

\section{Phenomenological fit}
\label{app:Phenomenological_fit}

The phenomenological template factors out the pieces whose scaling is already
understood analytically and fits only the smooth remainder. In the scanned small-$\alpha_0$ regime, $z^2(x_f)\propto\alpha_0$ motivates an explicit $\alpha_0^{-1}$ factor. Residual $\alpha_0$ dependence of the evolved mode is included in the fit error.
The dimensional factor $(\tau_B^2\Mpl^2)^{-1}$ is also kept explicit, since, at
fixed dimensionless $\kappa$, it fixes the overall energy scale. The remaining
dependence on $d$ and $\kappa$ is described by an effective growth function,
which approximates the accumulated evolution from the oscillatory regime to the finite amplitude at the end of Genesis.

The resulting fit is
\begin{equation}
    \mathcal{P}_\mathcal{R}^{\text{fit}}
    =
    \frac{1}{4\pi^2\tau_B^2\Mpl^2\alpha_0}
    \exp\!\left[
        -\frac{2(1+d)}{d}
        +
        \frac{2J(d,\kappa)}{d}
        +
        c_0+c_1d
    \right]
    C_{\nu_\star}^{\,q}\,.
    \label{eq:PR_fit}
\end{equation}
Here
\begin{equation}
    C_\nu
    =
    2^{2\nu-3}
    \left[
        \frac{\Gamma(\nu)}{\Gamma(3/2)}
    \right]^2\,,
\end{equation}
and we define
\begin{equation}
    A(d)=3(1+d)^2\,,
    \qquad
    \xi(d,\kappa)=\kappa^d\,.
\end{equation}
The effective index is
\begin{equation}
    \nu_\star
    =
    \left[
        \frac94
        +
        A\,\xi(1-\xi)
    \right]^{1/2}\,,
\end{equation}
and the auxiliary growth variable is
\begin{equation}
    u_\star
    =
    \frac{A(1-\xi)}{\nu_\star+3/2}\,.
\end{equation}
The function \(J(d,\kappa)\) is given by
\begin{equation}
    J(d,\kappa)
    =
    3
    +
    \sqrt{A}\arctan\!\left(\frac{u_\star}{\sqrt A}\right)
    +
    \frac32
    \ln\!\left(\frac{A}{A+u_\star^2}\right)
    -
    \frac{A(3+u_\star)}{A+u_\star^2}\,.
\end{equation}
In the formal template limit $\kappa^d\to0$,
\begin{equation}
    \nu_\star\to\frac32\,,
    \qquad
    C_{\nu_\star}\to1\,.
\end{equation}
However, \(J(d,\kappa)\) does not vanish in this limit. Instead,
\begin{equation}
    J(d,\kappa)\to J_0(d)\,,
\end{equation}
where
\begin{equation}
    J_0(d)
    =
    \sqrt{3}(1+d)
    \arctan\!\left(\frac{1+d}{\sqrt3}\right)
    +
    \frac32
    \ln\!\left[
        \frac{3}{3+(1+d)^2}
    \right]\,.
\end{equation}
The template therefore has a formally scale-independent infrared limit, but this should not be extrapolated beyond the scan.
Only part of the scan is nearly scale-invariant. For example, $d=0.1$, $\alpha_0=10^{-23}$, and $\kappa=10^{-4}$ give $n_s\simeq0.48958$ from the full mode equation. The template fits amplitudes across the full scan and does not imply near scale invariance everywhere.
The \(d\)-dependent constant \(J_0(d)\) contributes only
to the overall amplitude. The factor \(C_{\nu_\star}\) has the form of the
usual Hankel-amplitude factor, while the exponent \(q\) parametrizes the
residual departure from a pure Hankel approximation.

The numerical coefficients are
\begin{subequations}
    \begin{align}
        c_0 & = 0.0139557668608\,, \\
        c_1 & = -0.275286738213\,, \\
        q & = 1.32186006083\,.
    \end{align}
\end{subequations}
They are obtained by a fit that minimizes the maximum logarithmic error.
Explicitly, after removing the known contribution
\begin{equation}
    \ln \mathcal{P}_\mathcal{R}^{\text{base}}
    =
    -\ln\!\left(4\pi^2\tau_B^2\Mpl^2\alpha_0\right)
    -
    \frac{2(1+d)}{d}
    +
    \frac{2J(d,\kappa)}{d}\,,
\end{equation}
we fit the residual
\begin{equation}
    y_i
    =
    \ln \mathcal{P}_{\mathcal{R},i}^{\text{num}}
    -
    \ln \mathcal{P}_{\mathcal{R},i}^{\text{base}}
\end{equation}
by
\begin{equation}
    y_i\simeq c_0+c_1d_i+q\ln C_{\nu_{\star,i}}\,.
\end{equation}
The parameters are fixed by minimizing the maximal absolute logarithmic error,
\begin{equation}
    \min_{c_0,c_1,q,t} \, t\,,
    \qquad
    \left|
        c_0+c_1d_i+q\ln C_{\nu_{\star,i}}-y_i
    \right|
    \leq t\,,
\end{equation}
which is a linear-programming problem once the template functions are fixed.

The numerical scan uses an \(80\times80\times80\) grid,
\begin{equation}
    10^{-30}\leq \alpha_0\leq 10^{-21}\,,
    \qquad
    0.1\leq d\leq 0.4\,,
    \qquad
    10^{-9}\leq \kappa\leq 10^{-4}\,.
\end{equation}
In the data file, the column denoted by \texttt{alpha} corresponds to
\(\alpha_0\). The fit should not be extrapolated outside this range.
The precomputed numerical table, fitting notebook, Wolfram notebook, and
validation plots are available in the companion GitHub repository\footnote{
\href{https://github.com/Furton/Genesis-in-Minimally-Modified-Gravity}
{github.com/Furton/Genesis-in-Minimally-Modified-Gravity}.
}. The table stored in the repository should be interpreted as the
dimensionless quantity $\tau_B^2\Mpl^2\mathcal{P}_\mathcal{R}$; the physical amplitude follows from the explicit factor $(\tau_B^2\Mpl^2)^{-1}$
in equation~\eqref{eq:PR_fit}.

The verification set holds out complete $(d,\kappa)$ pairs while retaining all $\alpha_0$ values. After removing the leading $\alpha_0^{-1}$ scaling, this mainly tests interpolation in $(d,\kappa)$. For the scalar amplitude, the mean relative error is about $0.11\%$, and the maximum remains below $0.28\%$ on both the training and verification sets.

\begin{table}
    \centering
    \scriptsize
    \resizebox{\linewidth}{!}{%
    \begin{tabular}{cccccc}
        \toprule
        $\alpha_0$ & $d$ & $\kappa$
        & $\mathcal{P}_\mathcal{R}^{\text{fit}}$
        & $\mathcal{P}_\mathcal{R}^{\text{num}}$
        & $\Delta_{\text{rel}}(\%)$
        \\
        \midrule
        $7.69265\times10^{-22}$ & $0.160759$ & $2.19671\times10^{-7}$
        & $1.4046197\times10^{16}$ & $1.4067648\times10^{16}$ & $0.152$
        \\
        $1.09138\times10^{-24}$ & $0.183544$ & $3.02700\times10^{-6}$
        & $2.2404214\times10^{19}$ & $2.2432384\times10^{19}$ & $0.126$
        \\
        $4.55227\times10^{-22}$ & $0.251899$ & $5.91769\times10^{-8}$
        & $7.0641840\times10^{17}$ & $7.0622356\times10^{17}$ & $0.0276$
        \\
        $2.26168\times10^{-25}$ & $0.164557$ & $6.27290\times10^{-6}$
        & $2.9927273\times10^{19}$ & $2.9948496\times10^{19}$ & $0.0709$
        \\
        \bottomrule
    \end{tabular}
    }
    \caption{
    Sample comparison between the phenomenological fit and the numerical
    spectrum. Here $\tau_B=-1$ and $\Mpl=1$, as in the dimensionless fitting convention, and
    \(\Delta_{\text{rel}}=100\,|\mathcal{P}_\mathcal{R}^{\text{fit}}/
    \mathcal{P}_\mathcal{R}^{\text{num}}-1|\).
    }
    \label{tab:fit_sample_points}
\end{table}

\begin{figure}
    \centering
    \begin{subfigure}{0.48\linewidth}
        \centering
        \includegraphics[width=\linewidth]{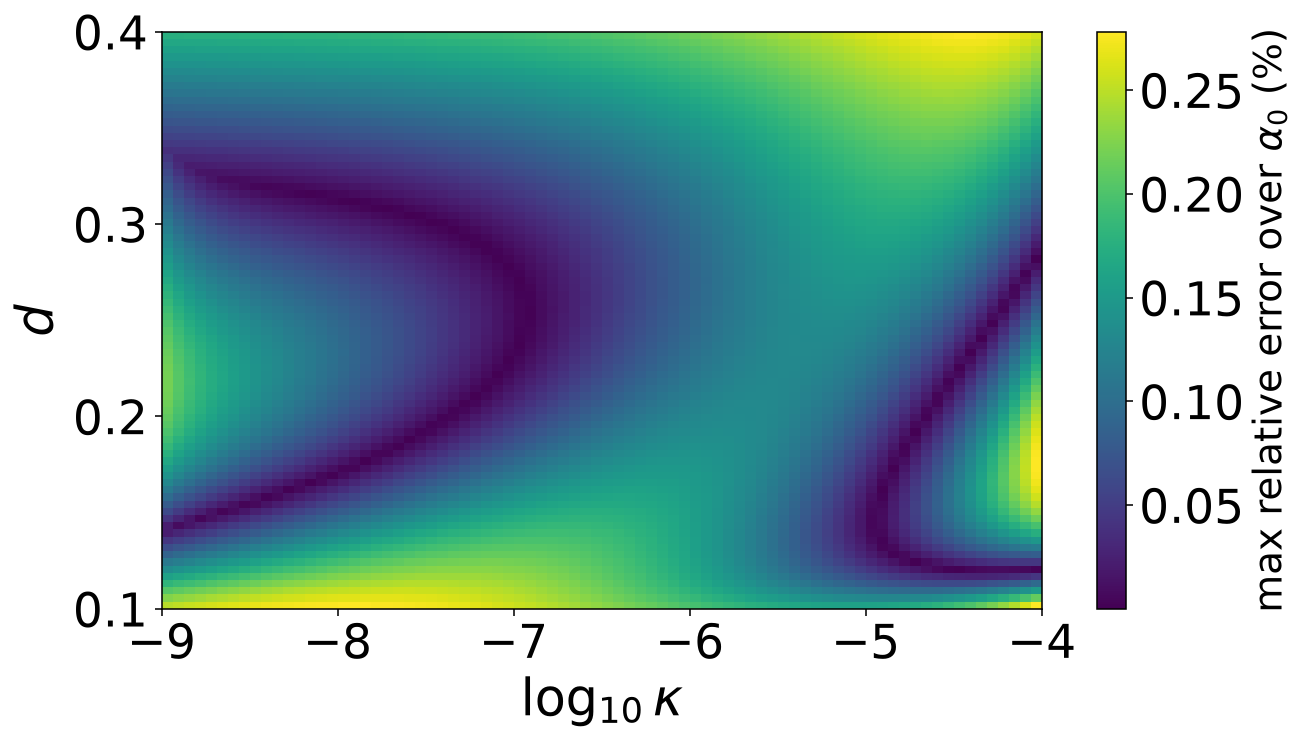}
        \caption{Relative error on the full grid.}
        \label{fig:fit_full_loss}
    \end{subfigure}
    \hfill
    \begin{subfigure}{0.48\linewidth}
        \centering
        \includegraphics[width=\linewidth]{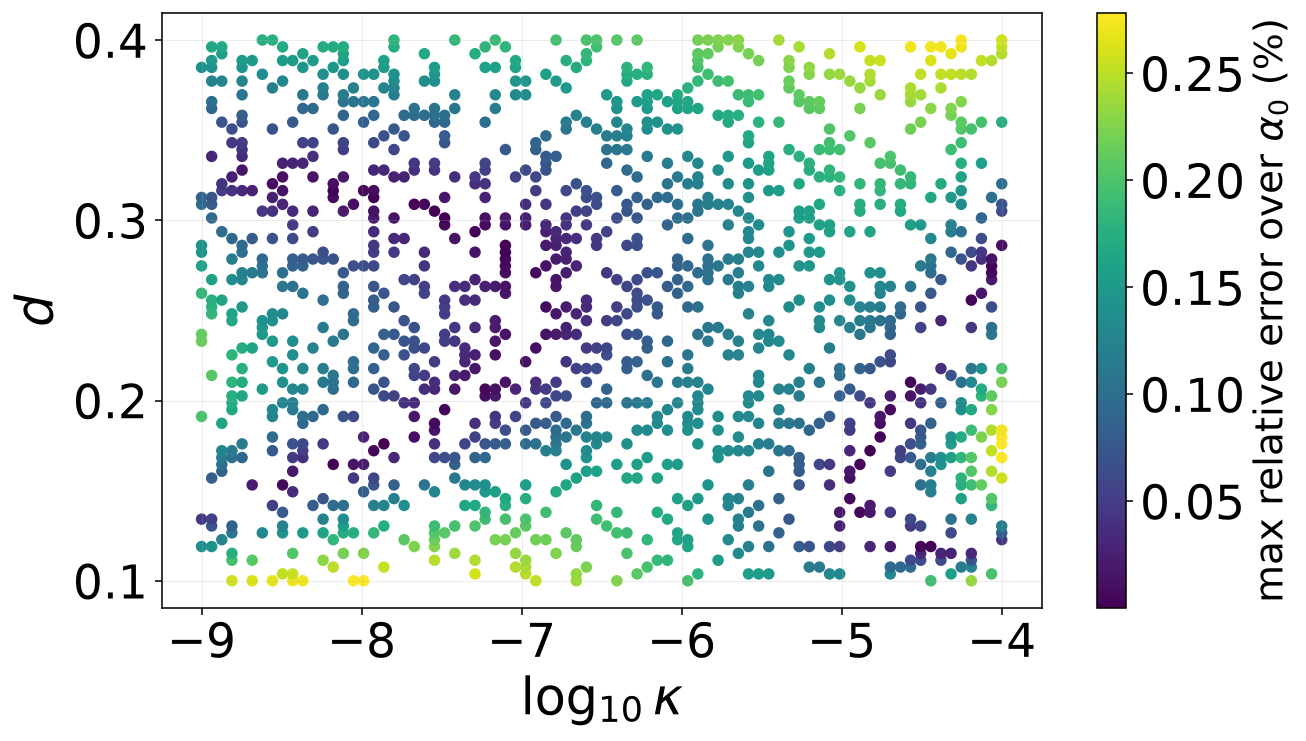}
        \caption{Relative error on the verification set.}
        \label{fig:fit_verification_loss}
    \end{subfigure}
    \caption{
    Accuracy of the phenomenological fit for the scalar power spectrum.
    The left panel shows the error over the full scan, while the right panel
    shows the held-out verification points. These tests indicate that the
    template interpolates the scanned \((d,\kappa)\) region accurately. They do
    not constitute a validation of extrapolations outside the scan range.
    }
    \label{fig:fit_validation}
\end{figure}

\section{Absence of asymptotic cubic strong coupling}
\label{app:Absence_Strong_Coupling}

\subsection{Reduced action and scalar perturbations}
\label{app:Quadratic_action_delta_chi}

The $\lambda_0$ equation derived from the VCDM action \eqref{action_mmg} gives
\begin{equation}
\lambda_0=-\frac23(K+\varphi)\,.
\end{equation}
Substituting this result back into the action gives
\begin{equation}
    S_{\text{VCDM}}
    =\frac{\Mpl^2}{2} \int \md^4x\,N\sqrt{\gamma}
    \left[
    {}^{(3)} R+K_{ij}K^{ij}-\frac13K^2
    +\frac43\varphi K+\frac23\varphi^2-2V(\varphi)
    -2 \lambda^i \frac{\partial_i \varphi}{N}\right]
    \,.
\end{equation}
At linear order, $\delta N$, $\beta$, and $\zeta$ are nondynamical. Varying them gives the constraints
\begin{equation}
    \delta N =\frac{\delta \chi'}{\bar\chi'}+ \frac{a^2U_{,\chi}}{\bar\chi'^2} \delta \chi +\Dsp\zeta\,,
\quad
\beta=-\frac{3\bar\chi'}{2\Mpl^2}\Delta^{-1}\delta \chi\,,\quad
    \zeta = \frac{(3-\Dsp)\frac{\delta\chi'}{\bar\chi'}-(3+\Dsp) \frac{a^2U_{,\chi}}{\bar\chi'^2} \delta \chi}{\Dsp(1+\Dsp)}\,,
    \label{eq:Constraints_for_delta_chi}
\end{equation}
where
\begin{equation}
  \Dsp\equiv\frac{2\Mpl^2}{\bar\chi'^2}\Delta\,.
\end{equation}
At $\Dsp=-1$, the algebraic constraint solution at fixed $(\delta\chi,\delta\chi^\prime)$ loses rank; consistency requires
\[
2\bar\chi' \delta\chi'=a^2U_{,\chi} \delta \chi\,.
\] For nonzero momentum, the canonical Hamiltonian system remains regular; this is a singularity of the velocity reduction, not of the physical linear evolution.
Using the constraint equations, the quadratic action reads in Fourier space
\begin{equation}
    S_{\delta\chi^2}^{(2)} = \frac{1}{2} \int \md\tau\int \md^3k\, a^2 K(k,\tau) \Big[ \vert \delta \chi^{\prime } \vert^2 - c_{\delta\chi}^2( k,\tau) k^2 \vert \delta \chi \vert^2 \Big]\,,
    \label{eq:2nd_order_delta_chi}
\end{equation}
where
\begin{align}
    K = &\frac{(\tilde  \kappa^2+3)^2}{\tilde  \kappa^2(\tilde  \kappa^2-1)}\,, \\
    c_{\delta \chi}^2 = & \frac{\tilde  \kappa^2 -1 }{3 + \tilde  \kappa^2  } - \frac{4 \Mpl^2 (1+x)^{-2d}}{(\tilde  \kappa^2 -1) (\tilde  \kappa^2 +3)  \tilde  \kappa^2 \chi_1^2 }\Big( - 2 (1+d)^2  (\tilde  \kappa^2 -1) \tilde  \kappa^2  + (1+d) (1+x)^d \nonumber \\
    & \times ( 6 +d (\tilde  \kappa^2 -3 ) (\tilde  \kappa^2 -1) + \tilde  \kappa^2 (\tilde  \kappa^2 -11) ) + (1+x)^{2d} ( 3 + \tilde  \kappa^2 (\tilde  \kappa^2 - 6)  )  \Big)\,.
\end{align}
In Fourier space, $\Dsp=-\tilde\kappa^2$ with $\tilde\kappa^2=2\Mpl^2k^2/\bar\chi^{\prime2}$. The velocity-reduced $\delta\chi$ action has a pole at
\begin{equation}
k_\star(x)=\frac{\chi_1 }{\sqrt{2}\Mpl (-\tau_B) (1+x) }\,.
\end{equation}
For $0<k<k_\star$, the velocity-reduced coefficient $K$ is negative. Because $\delta\chi$ mixes the curvature coordinate and momentum, this does not signal an additional ghost; the regular curvature or canonical phase-space description should be used, and this interval need not coincide with the region with $c_\mathcal{R}^2<0$ or the canonical non-oscillatory regime.
At fixed nonzero momentum, $x\to\infty$ implies $\tilde\kappa\to\infty$, and the leading terms are
\begin{equation}
    K \simeq 1 + \frac{7}{\tilde  \kappa^2 }\, \qquad c_{\delta \chi}^2 \simeq 1 - 4  \frac{\Mpl^2 + \chi_1^2}{\chi_1^2 \tilde  \kappa^2 }\,.
\end{equation}
Neglecting the sound-speed correction also requires $\tilde\kappa^2\gg4(1+\Mpl^2/\chi_1^2)$, while $\tilde\kappa\gg1$ controls only the kinetic correction, and subleading background terms must also be small.

\subsection{Cubic action}
\label{app:Cubic_action_Power_counting}

Let $\mathcal A=(n,\zeta,\beta)$, with $n\equiv\delta N$, denote the
scalar auxiliary fields. The transverse shift vanishes at first order,
$S_i^{(1)}=0$. We expand
$\mathcal A=\mathcal A^{(1)}+\mathcal A^{(2)}+\cdots$,
where $\mathcal A^{(n)}$ denotes the $n$th-order contribution, including
the factorial factor. Away from singular constraint surfaces,
\begin{align}
    S_{\text{red}}^{(3)}
    &=S_3[\mathcal A^{(1)}]
    +\left.\frac{\delta S_2}{\delta\mathcal A}\right|_{\mathcal A^{(1)}}
    \mathcal A^{(2)}
    \nonumber\\
    &=S_3[\mathcal A^{(1)}]\,.
\end{align}
Hence only the linear solutions are needed for the complete reduced cubic
action. After using the background
equations and discarding total derivatives, the complete cubic action for the scalar sector is
\begin{align}
    S_{\delta\chi^3}^{(3)}
        =\int\md\tau\,\md^3x\,a^2\Bigg\{\,
    & 2\Mpl^2\zeta\,\partial_i \delta N\,\partial_i\zeta
        +\Mpl^2(\delta N+\zeta)(\partial\zeta)^2
        +\frac{\Mpl^2}{8}(3\zeta-\delta N)B^i{}_jB^j{}_i
        \nonumber\\
    &+\frac12(3\zeta-\delta N)\delta\chi'^2
        -\delta\chi'\partial_i\beta\,\partial_i\delta\chi
        +\bar\chi^\prime \delta\chi'\left(\delta N^2-3\delta N\zeta+\frac92\zeta^2\right)
        \nonumber\\
    &-\bar\chi^\prime(3\zeta-\delta N)\partial_i\beta\,\partial_i\delta\chi
        +\frac{\bar\chi^{\prime 2}}{2}\delta N^2(3\zeta-\delta N)
        -\frac12(\delta N+\zeta)(\partial \delta\chi)^2
        \nonumber\\
    &-a^2\left[
        \frac16U_{,\chi\chi\chi}\delta\chi^3
        +\frac12U_{,\chi\chi}(\delta N+3\zeta)\delta\chi^2
        +U_{,\chi}\left(3\delta N\zeta+\frac92\zeta^2\right)\delta\chi
        \right]\Bigg\}\,,
    \label{eq:all_q3}
\end{align}
where
\begin{equation}
    B^i{}_j
    \equiv
    2\left(\partial^i\partial_j-\frac13\delta^i{}_j\Delta\right)\beta\,.
\end{equation}
The two-scalar/one-tensor terms are
\begin{align}
    S_{\delta\chi^2h}^{(3)}
    =\int\md\tau\,\md^3x\,a^2\Bigg\{\,&
    \frac12h^{ij}\partial_i\delta\chi\,\partial_j\delta\chi
    -\Mpl^2h^{ij}
    \left(2\partial_i \delta N\,\partial_j\zeta
    +\partial_i\zeta\,\partial_j\zeta\right)
    \nonumber\\
    &+\frac{\Mpl^2}{4}\left[
    -(3\zeta-\delta N)h'^{ij}B_{ij}
    +B^{ij}\partial^k\beta\,\partial_kh_{ij}
    \right]\Bigg\}\,.
    \label{eq:all_q2h}
\end{align}
The one-scalar/two-tensor terms are
\begin{align}
    S_{\delta\chi h^2}^{(3)}
    ={}&\frac{\Mpl^2}{2}
    \int\md\tau\,\md^3x\,a^2
    (\delta N+\zeta)\widetilde R^{(2)}[h]
    \nonumber\\
    &+\frac{\Mpl^2}{8}
    \int\md\tau\,\md^3x\,a^2
    \left[(3\zeta-\delta N)h'^i{}_jh'^j{}_i
    -2h'^i{}_j\partial^k\beta\,\partial_kh^j{}_i\right]\,,
    \label{eq:all_qh2}
\end{align}
where $\widetilde R^{(2)}[h]\equiv[{}^{(3)}R(e^h)]_{\mathcal O(h^2)}$
is the quadratic intrinsic-curvature contribution, a local sum of
$h\,\partial^2h$ and $(\partial h)^2$.
There is no direct matter $\delta\chi h^2$ vertex because $\det(e^h)=1$ and $\partial_i\bar\chi=0$.

For pure tensors the complete action reads
\begin{equation}
S_{h^3}^{(3)}
=\frac{\Mpl^2}{4}
\int\md\tau\,\md^3x\,a^2
\left(h_{ik}h_{j\ell}-\frac12h_{ij}h_{k\ell}\right)
\partial_k\partial_\ell h_{ij}\,,
\label{eq:all_h3}
\end{equation}
up to a spatial boundary term.

Equations \eqref{eq:all_q3}--\eqref{eq:all_h3} give the possible classes of cubic operators shown in table~\ref{tab:class_summary}, where $A\in\{n,\zeta,k\beta\}$ and $\partial$ denotes either a time or spatial derivative.
\begin{table}[H]
    \centering
    \caption{Possible classes of cubic operators.}
    \label{tab:class_summary}
    \begin{tabular}{cc}
        \toprule
        Sector & Cubic classes \\ \midrule
        \multirow{2}{*}{$\delta\chi^3$}
        & $\Mpl^2\partial^2A^3,\quad
        (\partial \delta\chi)^2A,\quad
        \bar \chi^\prime \,\partial \delta\chi\,A^2,\quad
        \bar\chi^{\prime 2}A^3,$ \\
        & $U_{,\chi\chi\chi}\delta\chi^3,\quad U_{,\chi\chi}A\delta\chi^2,\quad U_{,\chi}A^2\delta\chi$>
        \\[2.5mm]
        $\delta\chi^2 h$
        & $h(\partial \delta\chi)^2,\quad \Mpl^2h\,\partial^2A^2$
        \\[2.5mm]
        $\delta\chi h^2$
        & $\Mpl^2A\,\partial^2h^2$
        \\[2.5mm]
        $h^3$
        & $\Mpl^2h^2\partial^2h$ \\
        \bottomrule
    \end{tabular}
\end{table}
Using the constraints \eqref{eq:Constraints_for_delta_chi}, at fixed
nonzero momentum the deep-Genesis limit $x\to\infty$ gives
$|\Dsp|\gg1$. In this limit, $\delta N$, $\beta$, and $\zeta$ are
suppressed by $(\bar\chi^\prime\Dsp)^{-1}$:
\begin{align}
    \delta N \simeq & 4 (\bar \chi^\prime \Dsp)^{-1} \delta \chi^\prime - 2 a^2 \frac{U_{,\chi}}{\bar\chi^\prime} (\bar \chi^\prime \Dsp)^{-1} \delta \chi \simeq \mathcal{O}(x^{-1}) \delta \chi^\prime + \mathcal{O}(x^{-2}) \delta \chi\,, \\
    \beta   = & -3 (\bar\chi^\prime\Dsp)^{-1} \delta\chi  \simeq \mathcal{O}(x^{-1}) \delta\chi\,, \\
    \zeta \simeq & - (\bar \chi^\prime \Dsp)^{-1} \delta \chi^\prime - a^2 \frac{U_{,\chi}}{\bar\chi^\prime} (\bar \chi^\prime \Dsp)^{-1} \delta \chi \simeq \mathcal{O}(x^{-1}) \delta \chi^\prime + \mathcal{O}(x^{-2}) \delta \chi\,.
\end{align}
Thus, at fixed nonzero momentum, $\delta N$, $\beta$, and $\zeta$
vanish asymptotically. The leading pure-scalar cubic interaction is
$(\partial\delta\chi)^2A$, while the surviving interaction is Planck suppressed.

\bibliographystyle{JHEP}
\bibliography{refs}

\providecommand{\href}[2]{#2}\begingroup\raggedright\begin{thebibliography}{10}

\bibitem{Creminelli:2010ba}
P.~Creminelli, A.~Nicolis and E.~Trincherini, \emph{{Galilean Genesis: An Alternative to inflation}}, \href{https://doi.org/10.1088/1475-7516/2010/11/021}{\emph{JCAP} {\bfseries 11} (2010) 021} [\href{https://arxiv.org/abs/1007.0027}{{\ttfamily 1007.0027}}].

\bibitem{Creminelli:2012my}
P.~Creminelli, K.~Hinterbichler, J.~Khoury, A.~Nicolis and E.~Trincherini, \emph{{Subluminal Galilean Genesis}}, \href{https://doi.org/10.1007/JHEP02(2013)006}{\emph{JHEP} {\bfseries 02} (2013) 006} [\href{https://arxiv.org/abs/1209.3768}{{\ttfamily 1209.3768}}].

\bibitem{Creminelli:2006xe}
P.~Creminelli, M.A.~Luty, A.~Nicolis and L.~Senatore, \emph{{Starting the Universe: Stable Violation of the Null Energy Condition and Non-standard Cosmologies}}, \href{https://doi.org/10.1088/1126-6708/2006/12/080}{\emph{JHEP} {\bfseries 12} (2006) 080} [\href{https://arxiv.org/abs/hep-th/0606090}{{\ttfamily hep-th/0606090}}].

\bibitem{Rubakov:2014jja}
V.A.~Rubakov, \emph{{The Null Energy Condition and its violation}}, \href{https://doi.org/10.3367/UFNe.0184.201402b.0137}{\emph{Phys. Usp.} {\bfseries 57} (2014) 128} [\href{https://arxiv.org/abs/1401.4024}{{\ttfamily 1401.4024}}].

\bibitem{Nishi:2015pta}
S.~Nishi and T.~Kobayashi, \emph{{Generalized Galilean Genesis}}, \href{https://doi.org/10.1088/1475-7516/2015/03/057}{\emph{JCAP} {\bfseries 03} (2015) 057} [\href{https://arxiv.org/abs/1501.02553}{{\ttfamily 1501.02553}}].

\bibitem{Planck:2018jri}
{\scshape Planck} collaboration, \emph{{Planck 2018 results. X. Constraints on inflation}}, \href{https://doi.org/10.1051/0004-6361/201833887}{\emph{Astron. Astrophys.} {\bfseries 641} (2020) A10} [\href{https://arxiv.org/abs/1807.06211}{{\ttfamily 1807.06211}}].

\bibitem{Planck:2019kim}
{\scshape Planck} collaboration, \emph{{Planck 2018 results. IX. Constraints on primordial non-Gaussianity}}, \href{https://doi.org/10.1051/0004-6361/201935891}{\emph{Astron. Astrophys.} {\bfseries 641} (2020) A9} [\href{https://arxiv.org/abs/1905.05697}{{\ttfamily 1905.05697}}].

\bibitem{Libanov:2016kfc}
M.~Libanov, S.~Mironov and V.~Rubakov, \emph{{Generalized Galileons: instabilities of bouncing and Genesis cosmologies and modified Genesis}}, \href{https://doi.org/10.1088/1475-7516/2016/08/037}{\emph{JCAP} {\bfseries 08} (2016) 037} [\href{https://arxiv.org/abs/1605.05992}{{\ttfamily 1605.05992}}].

\bibitem{Kobayashi:2016xpl}
T.~Kobayashi, \emph{{Generic instabilities of nonsingular cosmologies in Horndeski theory: A no-go theorem}}, \href{https://doi.org/10.1103/PhysRevD.94.043511}{\emph{Phys. Rev. D} {\bfseries 94} (2016) 043511} [\href{https://arxiv.org/abs/1606.05831}{{\ttfamily 1606.05831}}].

\bibitem{Kolevatov:2017voe}
R.~Kolevatov, S.~Mironov, N.~Sukhov and V.~Volkova, \emph{{Cosmological bounce and Genesis beyond Horndeski}}, \href{https://doi.org/10.1088/1475-7516/2017/08/038}{\emph{JCAP} {\bfseries 08} (2017) 038} [\href{https://arxiv.org/abs/1705.06626}{{\ttfamily 1705.06626}}].

\bibitem{Mironov:2019qjt}
S.~Mironov, V.~Rubakov and V.~Volkova, \emph{{Genesis with general relativity asymptotics in beyond Horndeski theory}}, \href{https://doi.org/10.1103/PhysRevD.100.083521}{\emph{Phys. Rev. D} {\bfseries 100} (2019) 083521} [\href{https://arxiv.org/abs/1905.06249}{{\ttfamily 1905.06249}}].

\bibitem{Ilyas:2020zcb}
A.~Ilyas, M.~Zhu, Y.~Zheng and Y.-F.~Cai, \emph{{Emergent Universe and Genesis from the DHOST Cosmology}}, \href{https://doi.org/10.1007/JHEP01(2021)141}{\emph{JHEP} {\bfseries 01} (2021) 141} [\href{https://arxiv.org/abs/2009.10351}{{\ttfamily 2009.10351}}].

\bibitem{Zhu:2021ggm}
M.~Zhu and Y.~Zheng, \emph{{Improved DHOST Genesis}}, \href{https://doi.org/10.1007/JHEP11(2021)163}{\emph{JHEP} {\bfseries 11} (2021) 163} [\href{https://arxiv.org/abs/2109.05277}{{\ttfamily 2109.05277}}].

\bibitem{Ageeva:2020gti}
Y.~Ageeva, O.~Evseev, O.~Melichev and V.~Rubakov, \emph{{Toward evading the strong coupling problem in Horndeski genesis}}, \href{https://doi.org/10.1103/PhysRevD.102.023519}{\emph{Phys. Rev. D} {\bfseries 102} (2020) 023519} [\href{https://arxiv.org/abs/2003.01202}{{\ttfamily 2003.01202}}].

\bibitem{Ageeva:2020buc}
Y.~Ageeva, P.~Petrov and V.~Rubakov, \emph{{Horndeski genesis: consistency of classical theory}}, \href{https://doi.org/10.1007/JHEP12(2020)107}{\emph{JHEP} {\bfseries 12} (2020) 107} [\href{https://arxiv.org/abs/2009.05071}{{\ttfamily 2009.05071}}].

\bibitem{GilChoi:2025hbs}
H.~Gil~Choi, P.~Petrov and M.~Yamaguchi, \emph{{Can Horndeski Genesis be nonpathological?}}, \href{https://doi.org/10.1007/JHEP08(2025)044}{\emph{JHEP} {\bfseries 08} (2025) 044} [\href{https://arxiv.org/abs/2503.02626}{{\ttfamily 2503.02626}}].

\bibitem{DeFelice:2020eju}
A.~De~Felice, A.~Doll and S.~Mukohyama, \emph{{A theory of type-II minimally modified gravity}}, \href{https://doi.org/10.1088/1475-7516/2020/09/034}{\emph{JCAP} {\bfseries 09} (2020) 034} [\href{https://arxiv.org/abs/2004.12549}{{\ttfamily 2004.12549}}].

\bibitem{Afshordi:2006ad}
N.~Afshordi, D.J.H.~Chung and G.~Geshnizjani, \emph{{Cuscuton: A Causal Field Theory with an Infinite Speed of Sound}}, \href{https://doi.org/10.1103/PhysRevD.75.083513}{\emph{Phys. Rev. D} {\bfseries 75} (2007) 083513} [\href{https://arxiv.org/abs/hep-th/0609150}{{\ttfamily hep-th/0609150}}].

\bibitem{DeFelice:2022uxv}
A.~De~Felice, K.-i.~Maeda, S.~Mukohyama and M.C.~Pookkillath, \emph{{Comparison of two theories of Type-IIa minimally modified gravity}}, \href{https://doi.org/10.1103/PhysRevD.106.024028}{\emph{Phys. Rev. D} {\bfseries 106} (2022) 024028} [\href{https://arxiv.org/abs/2204.08294}{{\ttfamily 2204.08294}}].

\bibitem{DeFelice:2020cpt}
A.~De~Felice, S.~Mukohyama and M.C.~Pookkillath, \emph{{Addressing $H_0$ tension by means of VCDM}}, \href{https://doi.org/10.1016/j.physletb.2021.136201}{\emph{Phys. Lett. B} {\bfseries 816} (2021) 136201} [\href{https://arxiv.org/abs/2009.08718}{{\ttfamily 2009.08718}}].

\bibitem{Ganz:2022zgs}
A.~Ganz, P.~Martens, S.~Mukohyama and R.~Namba, \emph{{Bouncing cosmology in VCDM}}, \href{https://doi.org/10.1088/1475-7516/2023/04/060}{\emph{JCAP} {\bfseries 04} (2023) 060} [\href{https://arxiv.org/abs/2212.13561}{{\ttfamily 2212.13561}}].

\bibitem{Ganz:2024ihb}
A.~Ganz, P.~Martens, S.~Mukohyama and R.~Namba, \emph{{Bispectrum from inflation/bouncing Universe in VCDM}}, \href{https://doi.org/10.1142/S0217751X2540007X}{\emph{Int. J. Mod. Phys. A} {\bfseries 40} (2025) 2540007} [\href{https://arxiv.org/abs/2407.02882}{{\ttfamily 2407.02882}}].

\bibitem{Gumrukcuoglu:2016jbh}
A.E.~G{\"u}mr{\"u}k{\c{c}}{\"u}o{\u{g}}lu, S.~Mukohyama and T.P.~Sotiriou, \emph{{Low energy ghosts and the Jeans{\textquoteright} instability}}, \href{https://doi.org/10.1103/PhysRevD.94.064001}{\emph{Phys. Rev. D} {\bfseries 94} (2016) 064001} [\href{https://arxiv.org/abs/1606.00618}{{\ttfamily 1606.00618}}].

\bibitem{Ageeva:2018lko}
Y.A.~Ageeva, O.A.~Evseev, O.I.~Melichev and V.A.~Rubakov, \emph{{Horndeski Genesis: strong coupling and absence thereof}}, \href{https://doi.org/10.1051/epjconf/201819107010}{\emph{EPJ Web Conf.} {\bfseries 191} (2018) 07010} [\href{https://arxiv.org/abs/1810.00465}{{\ttfamily 1810.00465}}].

\bibitem{Ageeva:2021yik}
Y.~Ageeva, P.~Petrov and V.~Rubakov, \emph{{Nonsingular cosmological models with strong gravity in the past}}, \href{https://doi.org/10.1103/PhysRevD.104.063530}{\emph{Phys. Rev. D} {\bfseries 104} (2021) 063530} [\href{https://arxiv.org/abs/2104.13412}{{\ttfamily 2104.13412}}].

\bibitem{Cai:2022ori}
Y.~Cai, J.~Xu, S.~Zhao and S.~Zhou, \emph{{Perturbative unitarity and NEC violation in genesis cosmology}}, \href{https://doi.org/10.1007/JHEP10(2022)140}{\emph{JHEP} {\bfseries 10} (2022) 140} [\href{https://arxiv.org/abs/2207.11772}{{\ttfamily 2207.11772}}].

\bibitem{Akama:2022usl}
S.~Akama and S.~Hirano, \emph{{Primordial non-Gaussianity from Galilean Genesis without strong coupling problem}}, \href{https://doi.org/10.1103/PhysRevD.107.063504}{\emph{Phys. Rev. D} {\bfseries 107} (2023) 063504} [\href{https://arxiv.org/abs/2211.00388}{{\ttfamily 2211.00388}}].

\bibitem{Lyth:2005du}
D.H.~Lyth and Y.~Rodriguez, \emph{{Non-Gaussianity from the second-order cosmological perturbation}}, \href{https://doi.org/10.1103/PhysRevD.71.123508}{\emph{Phys. Rev. D} {\bfseries 71} (2005) 123508} [\href{https://arxiv.org/abs/astro-ph/0502578}{{\ttfamily astro-ph/0502578}}].

\end{thebibliography}\endgroup

\end{document}